\documentclass{aastex61}
\submitjournal{ApJ}
\usepackage{amsmath}
\usepackage{color}
\usepackage{natbib}
\begin{document}
\defcitealias{marble10}{M10}

\include{journals}

\shorttitle{Molecular cloud properties in N\,55}
\shortauthors{Naslim et al.}

\title{ALMA reveals molecular cloud N\,55 in the Large Magellanic Cloud as a site of massive star formation}

\correspondingauthor{Naslim Neelamkodan}
\email{nnks@uohyd.ernet.in}

\author[0000-0001-8901-7287]{Naslim~N}
\affiliation{Academia Sinica Institute of Astronomy and Astrophysics, Taipei 10617, Taiwan R.O.C}
\affiliation{Chile Observatory, National Astronomical Observatory of Japan, National Institutes of Natural Science, 2-21-1 Osawa, Mitaka, Tokyo 181-8588, Japan}
\affiliation{School of Physics, University of Hyderabad, Prof. C.Rao Road, Gachibowli, Telangana, Hyderabad, 500046, India}
\author{K.~Tokuda}
\affiliation{Department of Physical Science, Graduate School of Science, Osaka Prefecture University, 1-1 Gakuen-cho, Sakai, Osaka 599-8531, Japan}
\affiliation{Chile Observatory, National Astronomical Observatory of Japan, National Institutes of Natural Science, 2-21-1 Osawa, Mitaka, Tokyo 181-8588, Japan}

\author{T.~Onishi}
\affiliation{Department of Physical Science, Graduate School of Science, Osaka Prefecture University, 1-1 Gakuen-cho, Sakai, Osaka 599-8531, Japan}

\author[0000-0003-2743-8240]{F.~Kemper}
\affiliation{Academia Sinica Institute of Astronomy and Astrophysics, Taipei 10617, Taiwan R.O.C}
\author[0000-0002-7759-0585]{T.~Wong}
\affiliation{Department of Astronomy, University of Illinois, Urbana, IL 61801, USA}
\author{O.~Morata}
\affiliation{Academia Sinica Institute of Astronomy and Astrophysics, Taipei 10617, Taiwan R.O.C}

\author{S.~Takada}
\affiliation{Department of Physical Science, Graduate School of Science, Osaka Prefecture University, 1-1 Gakuen-cho, Sakai, Osaka 599-8531, Japan}

\author{R.~Harada}
\affiliation{Department of Physical Science, Graduate School of Science, Osaka Prefecture University, 1-1 Gakuen-cho, Sakai, Osaka 599-8531, Japan}

\author{A.~Kawamura}
\affiliation{Chile Observatory, National Astronomical Observatory of Japan, National Institutes of Natural Science, 2-21-1 Osawa, Mitaka, Tokyo 181-8588, Japan}

\author{K.~Saigo}
\affiliation{Chile Observatory, National Astronomical Observatory of Japan, National Institutes of Natural Science, 2-21-1 Osawa, Mitaka, Tokyo 181-8588, Japan}

\author{R.~Indebetouw}
\affiliation{Department of Astronomy, University of Virginia, PO Box 400325, VA 22904, USA}

\author{S.~C.~Madden}
\affiliation{Laboratoire AIM, CEA/DSM - CEA Saclay, 91191 Gif-sur-Yvette, France}

\author{S.~Hony}
\affiliation{Universit\"{a}t Heidelberg, Zentrum f\"{u}r Astronomie, Institut
f\"{u}r Theoretische Astrophysik, Albert-Ueberle-Str. 2, 69120 Heidelberg, Germany}
\author{M.~Meixner}
\affiliation{Space Telescope Science Institute, 3700 San Martin Drive, Baltimore, MD 21218, USA}

\begin{abstract}

We present the molecular cloud properties of N\,55 in the Large
Magellanic Cloud using $^{12}$CO(1-0) and $^{13}$CO(1-0) observations
obtained with Atacama Large Millimeter Array. We have done a detailed
study of molecular gas properties, to understand how the cloud
properties of N\,55 differ from Galactic clouds. Most CO emission
appears clumpy in N\,55, and molecular cores that have YSOs show
larger linewidths and masses. The massive clumps are associated with
high and intermediate mass YSOs. The clump masses are determined by
local thermodynamic equilibrium and virial analysis of the $^{12}$CO and $^{13}$CO
emissions. These mass estimates lead to the conclusion that, (a) the
clumps are in self-gravitational virial equilibrium, and (b) the
$^{12}$CO(1-0)-to-H$_2$ conversion factor, X$_{\rm CO}$, is
6.5$\times$10$^{20}$\,cm$^{-2}$\,(K\,km s$^{-1}$)$^{-1}$. This
CO-to-H$_2$ conversion factor for N55 clumps is measured at a spatial
scale of $\sim$0.67\,pc, which is about two times higher than the X$_{\rm CO}$
value of Orion cloud at a similar spatial scale. The core mass
function of N\,55 clearly show a turnover below 200\,M$_{\odot}$,
separating the low-mass end from the high-mass end. The low-mass end
of the $^{12}$CO mass spectrum is fitted with a power law of index
0.5$\pm$0.1, while for $^{13}$CO it is fitted with a power law index
0.6$\pm$0.2. In the high-mass end, the core mass spectrum is fitted
with a power index of 2.0$\pm$0.3 for $^{12}$CO, and with 2.5$\pm$0.4
for $^{13}$CO. This power-law behavior of the core mass function in
N\,55 is consistent with many Galactic clouds.

\end{abstract}
\section{Introduction}

Most stars form as clusters in Giant Molecular Clouds (GMCs) which
encompass cold molecular gas and dust with masses
$\sim$10$^{4-5}$M$_{\odot}$. Therefore, understanding the evolution of
dust and gas in GMCs is important to understand the formation of stars
in galaxies. The GMCs are composed of sub-parsec-sized clumps, the
size of which is determined by the forces of gravity and
magneto-turbulent pressure. Stars form inside these clumps, hence a
detailed understanding of the star formation process requires a
sub-parsec scale resolution view of GMCs and accurate measurements of
the physical parameters of these clumps. Star formation requires
high-density clumps where most of the interstellar hydrogen should be in
the form of molecular, H$_2$. Since H$_2$, in the form of cold
molecular gas, is almost totally undetectable with observations,
emission from the tracer CO and its isotopes is widely used in
galaxies to estimate the molecular core properties and distribution of
GMCs. There have been large-scale molecular surveys undertaken in
$^{12}$CO(1-0) and $^{13}$CO(1-0) emission of the Galactic clouds to
unveil distribution, structure and physical properties of molecular
clumps \citep{dame87, combes91, mizuno95, onishi96}. Infrared
observations of these molecular clouds have revealed young stellar objects (YSOs) embedded in dense molecular cores, that strongly suggest
on-going star formation. $^{13}$CO emission often can
be optically thin and traces dense molecular gas, therefore it can be
used to study the relationship between YSOs and their parent molecular
cloud \citep{bally89, fukui91}. Similar observations at high spatial
resolution in Galactic clouds have been extended to much denser
molecular tracers, such as NH$_3$, CS, HCN, and HCO$^+$ \citep{myers87, zhou89, myers93, Stutzki90}. 
These authors found that small dense cores, with size $\sim$ 0.1\,pc and density
n$\sim$10$^{4}$\,cm$^{-3}$ are the sites for massive star formation, and physical
conditions of cores are closely related to the properties of embedded
YSOs.

There is an extensive literature available on determining the physical
parameters of clumps in Galactic molecular clouds and their
relationships to star formation. Many of these studies focus on GMC
scaling relations between observable quantities such as size, velocity
dispersion, mass surface density, etc. \citet{larson81} reported three
empirical scaling relations for clump basic parameters in Milky Way
GMCs. Those are: 1) a power law relation between velocity dispersion
and size of emitting medium, $\sigma_v \propto R^{0.38}$; 2) molecular
clouds are virialized, $2\sigma_v R^2/GM\sim1$;
3) mean density of cloud is inversely related to size, $n \sim R^{-1.1}$.
The Larson scaling relationships have been later tested
with observations in various molecular tracers including rotational
transitions of CO, its isotopes and denser molecular tracers such as
NH$_3$, CS, HCN, and HCO$^+$ \citep{dame86, Scoville86, solomon87, zhou89,
  Hobson92, myers93, tatematsu98, heyer01, heyer09, ikeda09a}. Many of
these studies reveal power law relationships between the velocity
dispersion and size indicating self-gravitating clouds in virial
equilibrium with a power law index ranging from 0.25 to
0.75. In contrast, \cite{oka01} reported large velocity dispersion in the Galactic
center clouds with $^{12}$CO(1-0) observations, showing a deviation
from a power law relation between velocity dispersion and
radius. Their studies indicate the requirement of external pressure
for cloud equilibrium confinement. Spatially resolved
studies of molecular gas properties in the Large Magellanic Cloud (LMC), as well as certain dwarf galaxies and local group spiral
galaxies, show similar GMC characteristics as those found in the Milky
Way \citep{bolatto08, fukui08, hughes10, wong11}. The GMCs in these
galaxies show similar size, linewidths and masses as in Galactic
clouds. However, observation of
CO lines in metal-poor dwarf galaxies
indicates extremely weak CO luminosity compared to their star
formation rates. The CO luminosity to star formation rate ratio tends
to decrease with decreasing metallicity, although the dwarf galaxies may still follow the Kennicutt-Schmidt relationship depending on how the H$_2$ mass is accounted for \citep{Ohta93, Taylor98,
  Schruba12, cormier14}. Observations suggest that in metal-poor galaxies,
the clump properties and hence star formation process may differ from those
in metal-rich environments due to hard radiation field from newly formed stars. Due to
reduced dust shielding and molecular gas abundances, radiation can
penetrate more deeply into the molecular cloud and may produce a
clumpier cloud \citep{madden06, remy-ruyer13}. 

Spatially resolved CO observations in large-scale surveys have been
carried out in the nearest low-metallicity galaxy, the LMC, with an
aim to investigate whether the GMC characteristics and star formation
conditions follow universal patterns. The LMC is an excellent site to 
explore the GMC characteristics in a sub-solar
metallicity (0.5\,Z$_{\odot}$) due to its proximity (50\,kpc) and
favorable viewing angle \citep{Pietrzy13, van01}. The clumpy structures have
been revealed in CO observations of the LMC molecular clouds. For
example, the observations of $^{12}$CO and $^{13}$CO in J=1-0, 2-1,
3-2 and 4-3 transitions have revealed the dense and hot molecular
clumps, along with detailed physical properties of the molecular gas at a
spatial resolution of $\sim$5\,pc in H\,{\sc ii} regions \citep{minamid08,
  minamid11}. These observations were done at an angular resolution\,$\sim$\,23$^{\prime\prime}$, 
  which cannot resolve the clump sub-structures at
the LMC distance, 50\,kpc \citep{Pietrzy13}. The large-scale CO surveys in the LMC molecular
clouds include: 12$^{\prime}$ resolution map of $^{12}$CO(1-0) with
1.2\,m Columbia Millimeter-Wave Telescope; 2.6$^{\prime}$ resolution
map of $^{12}$CO(1-0) with the NANTEN 4\,m telescope; 45$^{\prime\prime}$
resolution $^{12}$CO(1-0) targeted mapping of known CO clouds from
NANTEN survey with the Mopra telescope (MAGMA \citet{wong11}); 1$^{\prime}$ resolution
individual cloud mapping in $^{12}$CO(1-0) with the Swedish-ESO
Submillimetre Telescope (SEST). \citet{fukui08} report the
CO-to-H$_2$ conversion factor to be 7$\times$10$^{20}$\,\rm cm$^{-2}$(K\,km\,s$^{-1}$)$^{-1}$ for the molecular clouds in the LMC from the NANTEN
survey at angular resolution 2.6$^{\prime}$. The size-linewidth relation is found to be consistent with the power law relation
proposed by \citet{larson81} with power law index of 0.5. \citet{wong11}
report a much steeper power law relation for size and velocity
dispersion. They suggest that this power law relation may break down when
the structures are decomposed into smaller structures.  

In addition, there has been high spatial resolution (sub-parsec)
mapping of $^{12}$CO(2-1) and $^{13}$CO(2-1) observations with the
Atacama Large Millimeter Array (ALMA) in the active star-forming
regions, 30\,Doradus and N\,159 west, of the LMC \citep{indebetow13,
  fukui15}. \citet{indebetow13} report relatively high velocity
dispersions for 30\,Doradus clouds and a power law relation
inconsistent with those found for Galactic clouds. They suggest the
large velocity dispersions for 30\,Doradus clouds must be due to
pressure confinement where an external force is necessary to hold
clumps in equilibrium. Furthermore, the dust mass is found to be a
factor of 2 lower than the average value of the LMC, indicating a
reduced dust-to-gas mass ratio in 30\,Doradus. \citet{fukui15}
  identified protostellar outflows toward two young high-mass stars
  along with an indication of colliding filaments in N\,159 west using
  $^{13}$CO(2-1) mapping at sub-parsec scale resolution. These
observations necessitate a further study of GMC properties in an LMC
molecular cloud, where the star formation activity should be
comparable to many other Milky Way clouds abiding Larson's scaling
relations.

This paper examines the molecular clump properties such as size,
velocity dispersion, mass, and their association to
star formation in N\,55 region of the LMC. This region is less extreme than
the well studied 30\,Doradus or N\,159 environments. We choose this cloud based on our
infrared spectroscopic and photometric observations with {\it Spitzer}
and {\it Hershel space telescopes} as part of Surveying the Agents of
Galaxy Evolution (SAGE) and the {\it Herschel} Inventory of the Agents
of Galaxy Evolution (HERITAGE) of the LMC \citep{meixner06,
  meixner10}. With infrared spectrograph on {\it Spitzer} (as part of
SAGE), we have detected the H$_2$ rotational transitions at 28.2 and
17.1${\,\rm \mu m}$ \citep{nas15} in N55. The clumpy and filamentary
structures of H$_2$ emission in N\,55 spatially resemble the
distribution of polycyclic aromatic hydrocarbon (PAH) emission traced
by {\it Spitzer} InfraRed Array Camera (IRAC) 8.0${\,\rm \mu m}$ as
well as the dust emission traced by {\it Herschel} Photodetector Array
Camera and Spectrometer (PACS) 100${\,\rm \mu m}$. A detailed analysis
of the infrared observations of this region will be presented in a
future paper. In this paper, we aim to determine how (dis-)similar the N\,55 clumps are from Milky Way clumps and
30\,Doradus. We study the molecular clump scaling
relations of N\,55 and investigate whether the power law relations for
Milky Way clouds hold in a sub-solar metallicity environment.

\section{N\,55 molecular cloud}

\begin{figure}
\centering
\includegraphics[scale=0.5, angle=0]{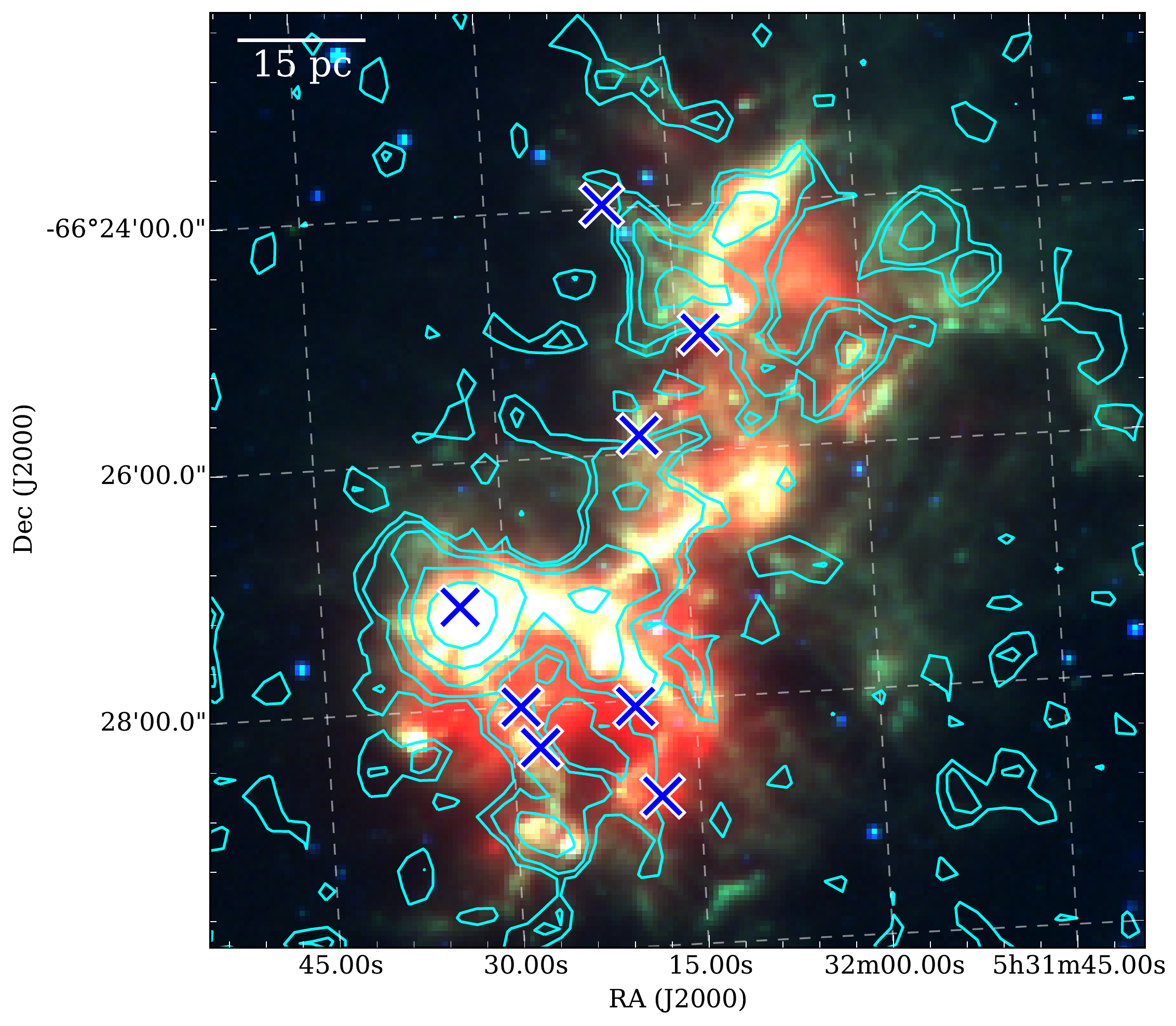}
\centering
\caption{The structure of N\,55 is shown in a composition of three color map: in blue, {\it Spitzer} IRAC 5.8${\,\rm \mu m}$ which traces infrared bright sources, in green {\it Spitzer} IRAC 8.0${\,\rm \mu m}$ which traces the PAH emission and in red MIPS 24${\,\rm \mu m}$ tracing hot dust heated by hard radiation from massive stars. The hot stars from \citet{olsen01} are labelled in crosses. For comparison, $^{12}$CO(3-2) emission observed with ASTE is shown in contours, tracing warm molecular gas.}
\label{3color}
\end{figure}

N\,55 is located inside LMC\,4, the largest Supergiant Shell (SGS) in
the LMC \citep{yamaguchi01}. The LMC\,4 has a diameter of about
1.5\,kpc. A study of the H${\alpha}$ map by \citet{olsen01} shows that
the H\,{\sc ii} region of N\,55 may be excited by a young cluster,
LH\,72, containing at least 8 O and B type stars. \citet{olsen01}
also reported the LMC 4 was not formed as a unit, but by overlapping
shells, and N\,55 seems to be associated with one of the overlapping
shells SGS\,14, which may have triggered the formation of
LH\,72. Therefore N\,55 seems to be located in an environment of
multiple supernova explosions, the most recent one being
SGS\,14. Complex filamentary structures are especially prominent
toward N\,55 in {\it Spitzer} SAGE images (Figure \ref{3color}), which
show the distribution of hot gas and photo-dissociated PAHs in the
interstellar medium. These filamentary structures may be due to
instabilities enhanced by shocks. Shocks of supernova explosions are
reported to be a cause of the enhanced filamentary structures
and dense clumps in molecular clouds \citep{ntormousi11}. The clumpy
nature of the molecular gas in N\,55 is revealed by $^{12}$CO(3-2)
emission observed with Atacama Submillimeter Telescope Experiment
(ASTE) (Figure \ref{3color}).  The interior of LMC\,4 is otherwise
almost empty of ionized and atomic gas, while N\,55 stands out as a
bright H\,{\sc ii} region in H${\alpha}$ map \citep{olsen01}. The
formation of LMC\,4 and the survival of N\,55 have been debated for a
decade.  Suggestions include supernova explosions, gamma ray bursts,
stellar winds from massive stars in LH\,72 or stochastic propagation
of star formation which have cleared out the gas from the interior
\citep{book09}. \citet{book09} suggest that star formation was likely
triggered by the compression of a pre-existing cloud in the expanding
shell, LMC\,4, as evidenced by the distribution and kinematics of the
gas in the expanding LMC\,4 \citep{israel03}.  Based on the
\emph{Spitzer} colors of point sources, \citet{gruendl09} and \citet{seale14}
identified 16 YSOs in N\,55 indicating on-going star formation.

\section{Observations}

N\,55 was observed with ALMA in cycle 1 \& 2 ( 2013.1.00214.S and
2012.1.00335.S), using band 3 receivers in spectral windows centered
at 115.27, 110.20 and 109.67 GHz with 70.557 kHz (0.2\,km\,s$^{-1}$)
spectral resolution. The area of coverage was
4$^{\prime}\times$6$^{\prime}$ at the center position: right ascension
$\sim 05^{\rm h}32^{\rm m}15^{\rm s}.49$ and declination $\sim
-66^{\rm o}26^{\prime}14^{\prime\prime}.00$. In each spectral window,
the correlator was set to have a bandwidth of 117.18 MHz. Uranus and
Ganymede were used as flux calibrators. The data were processed in the
Common Astronomy Software Application
(CASA\footnote{http://casa.nrao.edu}) package and visibilities
imaged. The synthesized beam for $^{12}$CO(1-0) is approximately
3.5$^{\prime\prime}\times$2.3$^{\prime\prime}$ which corresponds to 0.84$\times$0.55\,pc$^2$.
The achieved rms per channel over 0.2\,km\,s$^{-1}$
  is about $\sim$70 mJy\,beam$^{-1}$ (0.8\,K). For
  $^{13}$CO(1-0) the synthesized beam is about
  3.1$^{\prime\prime}\times$2.5$^{\prime\prime}$ which corresponds to
  0.74$\times$0.60\,pc$^2$, and the achieved
  sensitivity per channel over 0.2\,km\,s$^{-1}$ is
  20\,mJy\,beam$^{-1}$ (0.30\,K). Note that, our $^{13}$CO(1-0) data are three times more sensitive than the $^{12}$CO(1-0)
  data, due to the longer integration time and four repeated observations of
  the $^{13}$CO. The details of observations are given in Table \ref{obs}. The typical system noise temperatures of the $^{12}$CO and the
  $^{13}$CO are $\sim$170\,K and $\sim$90\,K. We use a
  moment masked cube \citep{dame11} to suppress the noise effect in
  our analysis. The moment masked cube has zero
  values at the emission free pixels, which is useful to avoid a large
  error arising from the random noise. The emission-free pixels are
  determined by identifying significant emission from the smoothed
  data whose noise level is much lower than the original data. The
  caveat of this method is that it eliminates small clouds with low peak
  intensity because we smooth the data both for the velocity and
  spatial directions \citep{dame11}. In order to determine the
fraction of missing flux caused by interferometric observation with
ALMA 12\,m array, we compare our data with the single dish Mopra
observation of $^{12}$CO(1-0). About 30$\%$ of the ALMA flux is
missing when smoothed to 45" and compared to the Mopra observations.

\begin{table}
\centering
\caption{Observations}
\label{obs}
\begin{tabular}{@{}ccccccc}
\hline
  & \multicolumn{3}{c}{$^{12}$CO(1-0)} &   \multicolumn{3}{c}{$^{13}$CO(1-0)}  \\
&      Execution Block       &    Date  & Integration time    &    Execution Block        &    Date  &  Integration time  \\
&                  &      & (minutes)    &            &     &  (minutes)  \\
\hline
&     uid://A002/X98ed3f/X1cf5             &  2015-01-06    & 70  &   uid://A002/X77da97/X177  & 2013-12-31 & 55 \\  
  &                        &               &               &  uid://A002/X77da97/X836    &  2014-01-01  & 51 \\            %   \\
  &                   &         &          &  uid://A002/X98ed3f/X21cf &   2015-01-06  & 65  \\
&                                      &            &                    &  uid://A002/X9a24bb/X1460  &   2015-01-22  & 57     \\               %\\
%  &   &     &      &   &    \\

\hline 

\end{tabular}
%\parbox{100mm}{
%a: Execution Block
%}
\end{table}

%\section{$^{12}$CO and $^{13}$CO observations with ALMA}
\section{Molecular gas distribution in N\,55}
\begin{figure*}
\centering
\includegraphics[scale=0.35]{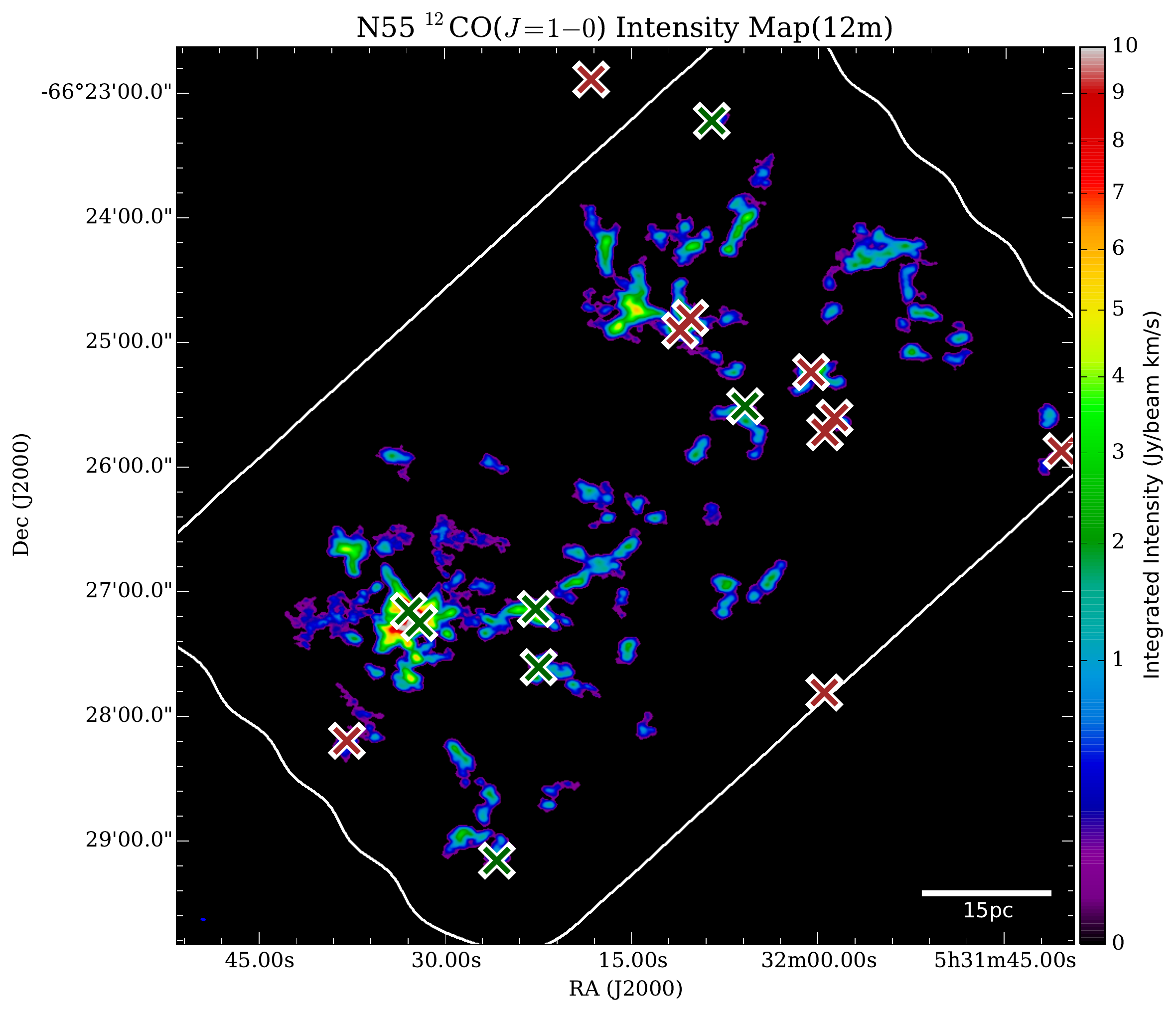}
\includegraphics[scale=0.35]{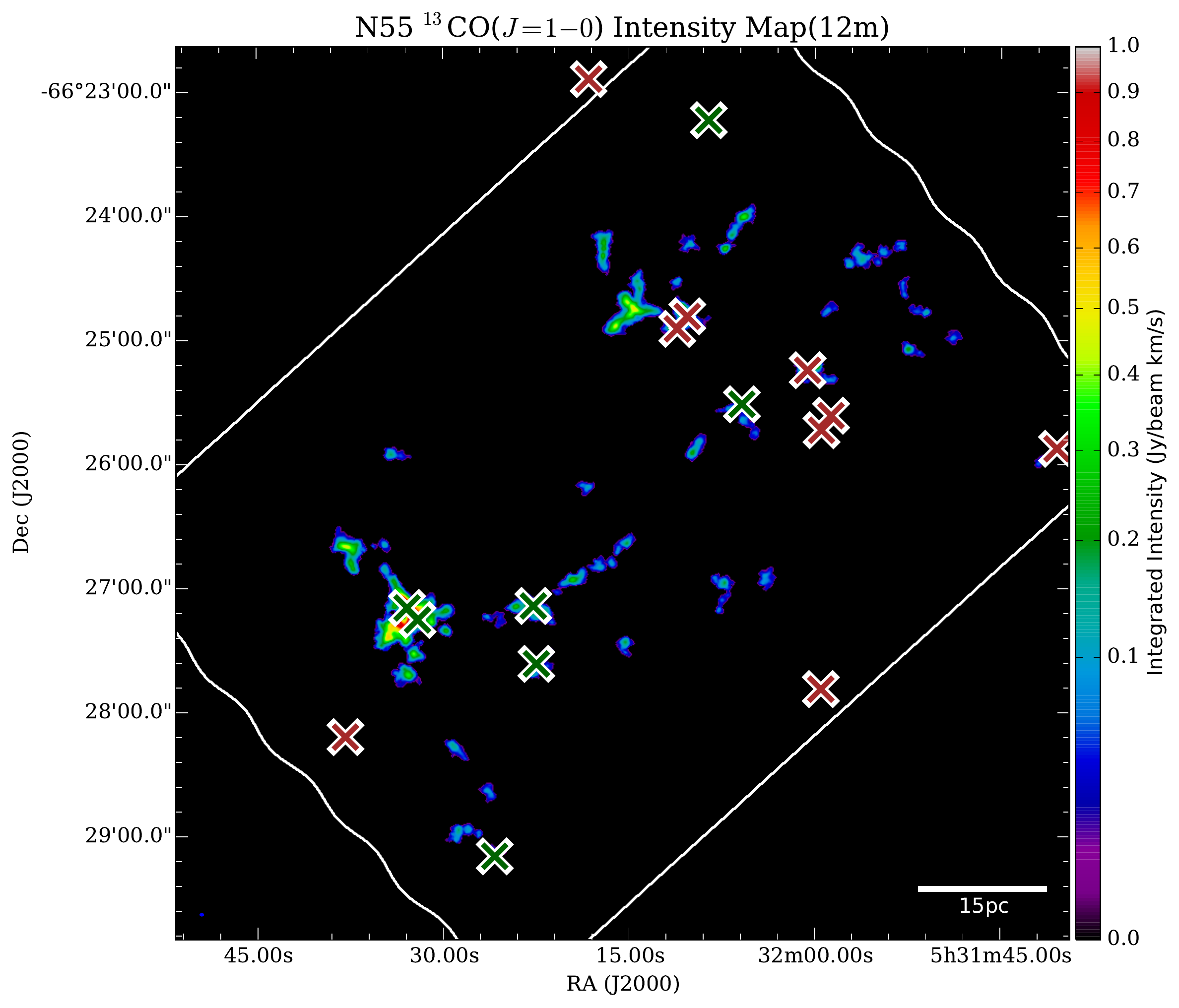}
\caption{$^{12}$CO(1-0) (left) and $^{13}$CO(1-0) (right) integrated intensity maps of N\,55. The YSOs identified with {\it Spitzer} (green-white crosses) and {\it Herschel} (red-white crosses) are labelled \citep{gruendl09, seale14}.}
\label{13CO}
\end{figure*}

\begin{figure*}
\centering
\includegraphics[scale=0.5]{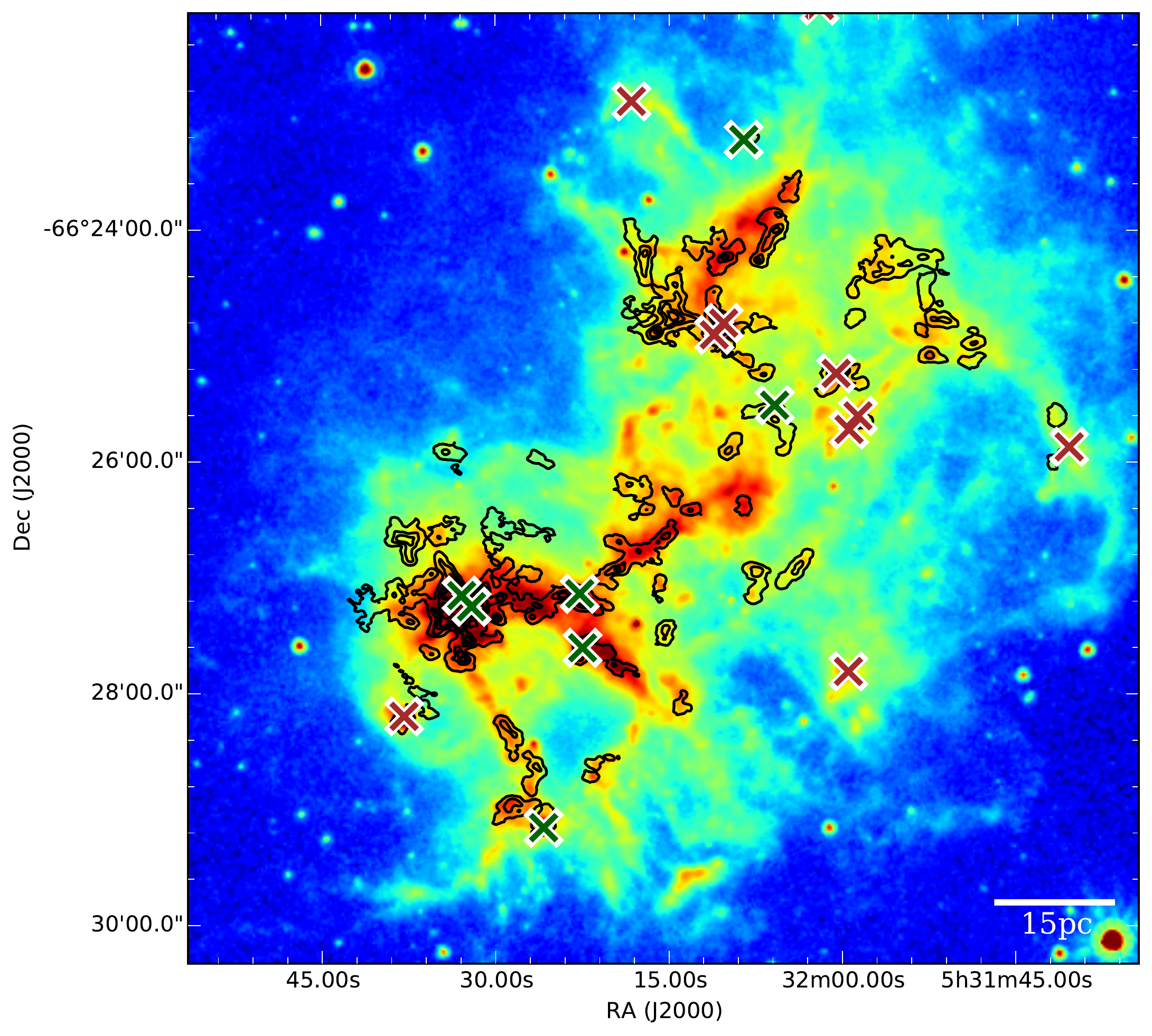}
\centering
\caption{$^{12}$CO(1-0) emission in contours on IRAC 8.0${\,\rm \mu m}$ map. The contour levels are 0.09, 1.5, 3.0, 4.5, 6.0, and 9.0 Jy\,beam$^{-1}$\,km\,s$^{-1}$. The YSOs identified with {\it Spitzer} (green-white crosses) and {\it Herschel} (red-white crosses) are labelled \citep{gruendl09, seale14}}
\label{CO-PAH}
\end{figure*}

Our $^{12}$CO(1-0) and $^{13}$CO(1-0) observations with ALMA show
the clumpy nature of molecular gas in N\,55 at sub-parsec scale (Figure \ref{13CO}). It is
clearly seen by eye that compact molecular clumps are distributed
along the filamentary structure of PAH emission traced by IRAC
8.0${\,\rm \mu m}$ (Figure \ref{CO-PAH}).
The $^{12}$CO(1-0) integrated intensity map appears to be
relatively extended, compared to that of its isotopologue $^{13}$CO(1-0) which
is clumpier and most of the clumps are clearly delineated by the
{\it Spitzer}-identified YSO candidates (Table \ref{YSO_clump}). These YSO positions are shown in
Figure \ref{13CO} along with $^{12}$CO(1-0) and $^{13}$CO(1-0)
integrated intensity maps. Many clouds show numerous sub-clumps and
these sub-clumps (molecular cores) are associated with YSOs. The
filamentary structures traced by $^{12}$CO(1-0) follows that of the
PAHs traced by IRAC 8.0${\,\rm \mu m}$ map
(Figure \ref{CO-PAH}). These observations may indicate that the CO is
mostly excited in the same regions where the dust is photoelectrically heated
by ultraviolet radiation. Since the structure is very complex, we identify the clumps
and determine their properties using a dendrogram method.

\section{Molecular mass determination}

\begin{figure*}
\centering
\includegraphics[scale=0.5]{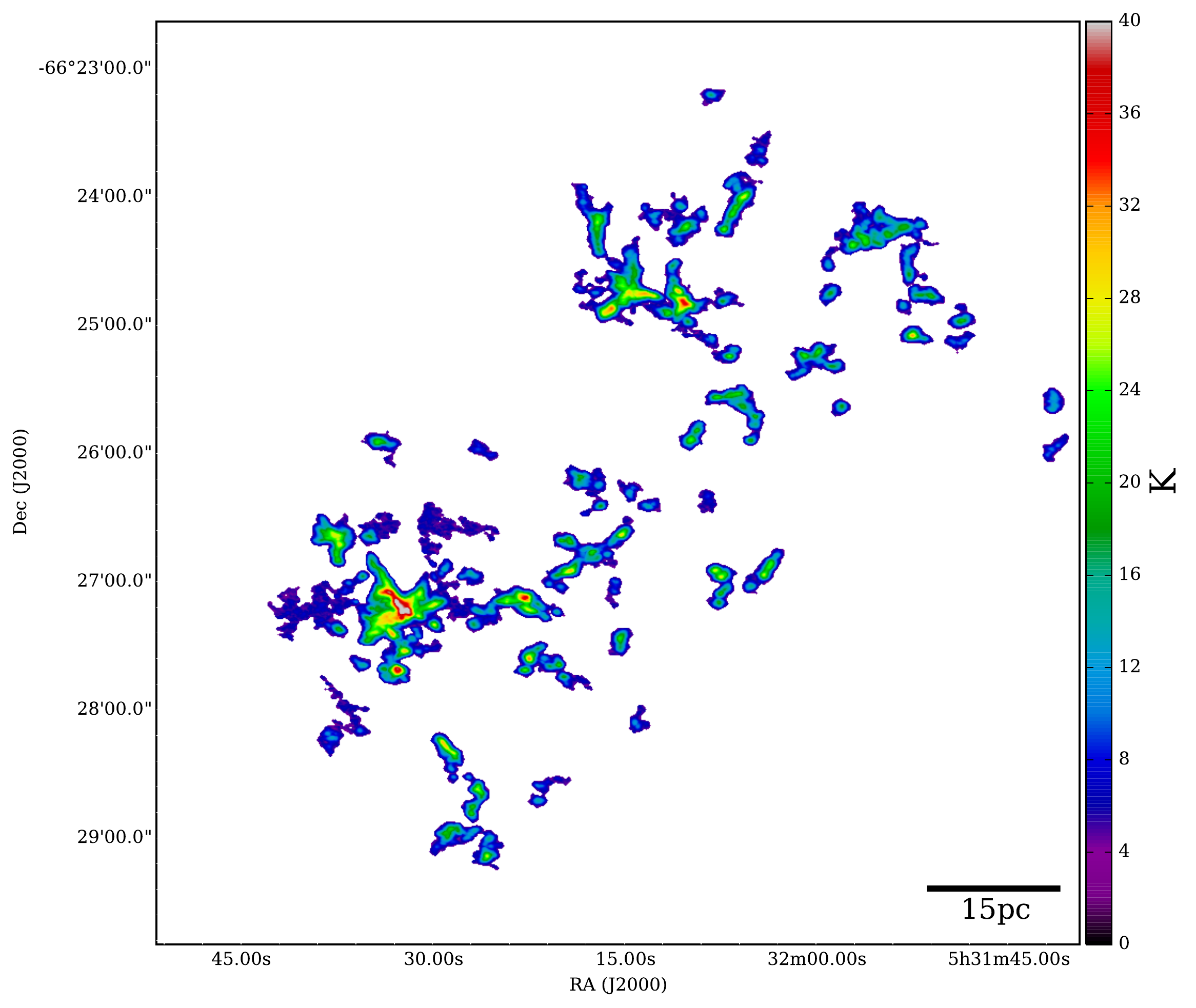}
\centering
\caption{Excitation temperature map of N\,55 which is derived from $^{12}$CO(1-0) emission using equation \ref{Text}.}
\label{text}
\end{figure*}

We determine the molecular gas mass using virial and local
thermodynamic equilibrium (LTE) assumptions. In order to determine the
LTE mass ($M_{\rm LTE}$) from $^{13}$CO(1-0) emission, we need to
calculate column density and assume that the excitation temperature for
$^{13}$CO(1-0) is the same as the $^{12}$CO(1-0) emission. To derive the
$^{13}$CO(1-0) column density we follow the method given by
\citet{dickman78}, \citet{pineda08}, \citet{pineda10} and \citet{nishimura15}. Assuming that
the emission is optically thin and in LTE, we derive the column density
using the following formula \citep{pineda08}.

\begin{equation}
    N(^{13}{\rm CO})=3.0\times 10^{14}\left(\frac{\tau(^{13}{\rm CO})}{1-e^{-\tau}}\right)\left(\frac{W({\rm CO})}{1-e^{-5.3/T_{{\rm ex}}}}\right)
    \label{column}
\end{equation}

In equation \ref{column}, $W$(CO) is the integrated intensity
($\sum{T_{{\rm mb}}}dv$) of $^{13}$CO(1-0) in K\,km\,s$^{-1}$.  To derive column
density of $^{13}$CO(1-0), we require the excitation temperature
$T_{\rm ex}$ and optical depth $\tau$. The excitation temperature $T_{\rm ex}$ at each position in 0.2\,km\,s$^{-1}$ wide channel is determined from the peak temperature
  ($T_{^{12}\rm CO}$) of the optically thick $^{12}$CO line.

\begin{equation}
     T_{{\rm ex}} = \frac{5.5\,{\rm K}}{{\rm ln}\{1+5.5\,{\rm K}/[T(^{12}{\rm CO})+0.82\,\rm {K}]\}}
    \label{Text}
\end{equation}

Using the excitation temperature derived from the above equation we calculate $^{13}$CO optical depth for each position in 0.2\,km\,s$^{-1}$ wide channel along the line of sight using the following equation:

\begin{equation}
\tau (^{13}{\rm CO}) = -{\rm ln}\left[1-\frac{T_{{\rm max}}/5.3}{1/({\rm e}^{5.3/T_{{\rm ex}}}-1)-0.16)}\right]
\end{equation}

Here $T_{\rm max}$ is the main beam brightness temperature at the peak of the $^{13}$CO emission.

The assumption of optically thin $^{13}$CO(1-0) emission and
  identical excitation temperatures for $^{13}$CO(1-0) and
  $^{12}$CO(1-0) in LTE, may cause several systematic errors
  in mass determination. In reality, the excitation condition for
  $^{12}$CO(1-0) and $^{13}$CO(1-0) is not the same throughout the
  cloud, as both these emission sample different volume
  densities. Moreover, the excitation of $^{13}$CO(1-0) is not fully
  thermalized in most of the cloud volume, however $^{12}$CO(1-0) can
  be thermalized even at densities lower than $^{12}$CO critical
  density ($n$H$_{2}<$750\,cm$^{3}$). \citet{heyer09} have quantitatively
  checked these effects by comparing the column densities derived by
  the LTE method with models obtained for several sets of cloud
  conditions using a large velocity gradient (LVG) approximation. They found that the LTE method can either
  underestimate or overestimate the input column density. In our study the typical densities of the cloud is $n$H$_{2}>1000$\,cm$^{-3}$. Based on Figure 6 in \citet{heyer09} the ratio of LTE
  to LVG model masses is around unity, however there is a scatter
  about 40$\%$.

We derive the total molecular gas mass using $^{13}$CO(1-0) column density and the molecular
abundance ratios [$^{12}$CO/$^{13}$CO] and [$^{12}$CO/H$_2$].
We adopt abundance ratios of 50 for [$^{12}$CO/$^{13}$CO] and
1.6$\times$10$^{-5}$ for [$^{12}$CO/H$_2$] which were derived for N\,159
in the LMC \citep{mizuno10}.
Hence, we obtain an H$_2$ column density of
$N({\rm H_2})=\frac{50}{1.6\times10^{-5}}N(^{13}\rm CO)$. Finally, the molecular
gas mass is calculated from $N({\rm H_2}$) at the LMC distance,
assuming that the mean molecular weight per H$_2$ is 2.7. 

We applied the above formula to molecular clumps of $^{13}$CO(1-0)
obtained with our clump decomposition method {\it astrodendro} (section 6)
and derived the LTE mass $M_{{\rm LTE}}$ for each clump, where the
excitation temperature of $^{13}$CO(1-0) is obtained from
the $T_{\rm ex}$ map (Figure \ref{text}) derived from equation \ref{Text}.

In addition to the LTE mass, we derive virial masses for
$^{13}$CO(1-0) and $^{12}$CO(1-0) clumps. We assume that clouds are
spherical with a truncated density profile, $\rho\sim$r$^{-1}$, and
derive the virial mass from the radius (R) and velocity dispersion
($\sigma_v$) of the clump \citep{wong11} as,

\begin{equation}
M_{\rm VIR} [M_{\odot}] = 1040\sigma_{v}^2 R
\end{equation}

\section{Clump decomposition using astrodendro}

The structure of molecular
clouds is highly hierarchical. Small scale dense molecular cores are
invariably enclosed within the envelope of large scale lower density
gas \citep{lada92}. These sub-parsec scale ($\sim$0.1\,pc) dense cores
are the sites for high mass star formation inside molecular
clouds. Probing the properties such as mass, radius and velocity
dispersion of these cores, allow us to understand the mass function as
well as conditions for star formation because of their close
relationship with newly formed stars \citep{difrancesco07}. In order to determine the properties of molecular clouds in
$^{13}$CO(1-0) and $^{12}$CO(1-0) emission, we use
the python package {\it astrodendro}. The
concept of dendrograms to characterize the structures of molecular
gas as a "structure tree" in a three-dimensional data cube was first
introduced by \citet{houslahan92}. A more systematic way of a
clump-finding algorithm using a dendrogram method was later implemented
by \citet{rosolowsky08}, who showed how to graphically
represent the hierarchical structure of nested isosurfaces in three-dimensional data cubes. 
A dendrogram is composed of leaves and branches where each entity in
the dendrogram is represented as an isosurface (three-dimensional
contour) in the data cube. The leaves are the brightest and smallest
structures that represent three-dimensional contours with single
local maxima. The local maxima represent the top level of the
dendrogram which are the brightest structures (dense molecular cores) that
we refer as leaves. Branches represent parent structures which
connect two leaves and form larger structures that represent lower
density media. Finally, all leaves and branches merge together to form
trunks of the tree. In this paper molecular cores are the
  structures which are identified as dendrogram {\it leaves}. Most of the leaves are of sub-parsec scale size. The clumps are those identified as dendrogram {\it trunks} which are mostly of parsec scale size. A cluster of clumps and cores is referred to as a cloud. With the use of dendrogram, we can represent the
complexity of molecular clouds effectively over a range of size scales
starting from low-density gas in the bottom level. 

{\it Astrodendro} identifies unique isosurfaces from each region of
emission in a three-dimensional data cube and computes the properties
based on the moment of volume weighted intensities of emission from
every pixel \citep{rosolowsky08}. The data cube consists of two
spatial dimensions (X and Y) and a velocity dimension (V). The rms
sizes \citep{colombo15} of a clump in two spatial dimensions ($\sigma_{\rm x}$ and
$\sigma_{\rm y}$), i.e. along the major axis and minor axis are
computed from the intensity-weighted second moments in two dimensions,
dx and dy. The geometric mean of these second spatial moment along
the major and minor axis of the clump is the final rms radius of a clump
($\sigma_r$). The rms radius in the second spatial moments obtained from {\it astrodendro}
is converted into the radius of a spherical cloud
$R=1.91\sigma_r$ \citep{solomon87}. 

The velocity dispersion, $\sigma_v$, is the intensity-weighted second
moment of the velocity axis. The sum of all emission within an
isosurface is the flux of a clump (F=$\sum{T_idxdydv}$), where $T_i$
is the brightness temperature. The luminosity of a clump is the integrated
flux scaled by the square of the distance to the object in parsec.

\begin{equation}
    L_{{\rm CO}} [{\rm K\,km\,s^{-1}\,pc^2}] = D^2 \sum(T) dxdydv
\end{equation}

To determine the uncertainties in the derived parameters, we use a
bootstrapping technique similar to the method given in
\citet{rosolowsky08}. We use 100 iterations to sample the derived
parameters. Tables \ref{t_lines1} and \ref{t_lines2} list
$^{12}$CO and $^{13}$CO clump (${\it astrodendro}$ trunk) properties.

\begin{figure}
\centering
\includegraphics[scale=0.5]{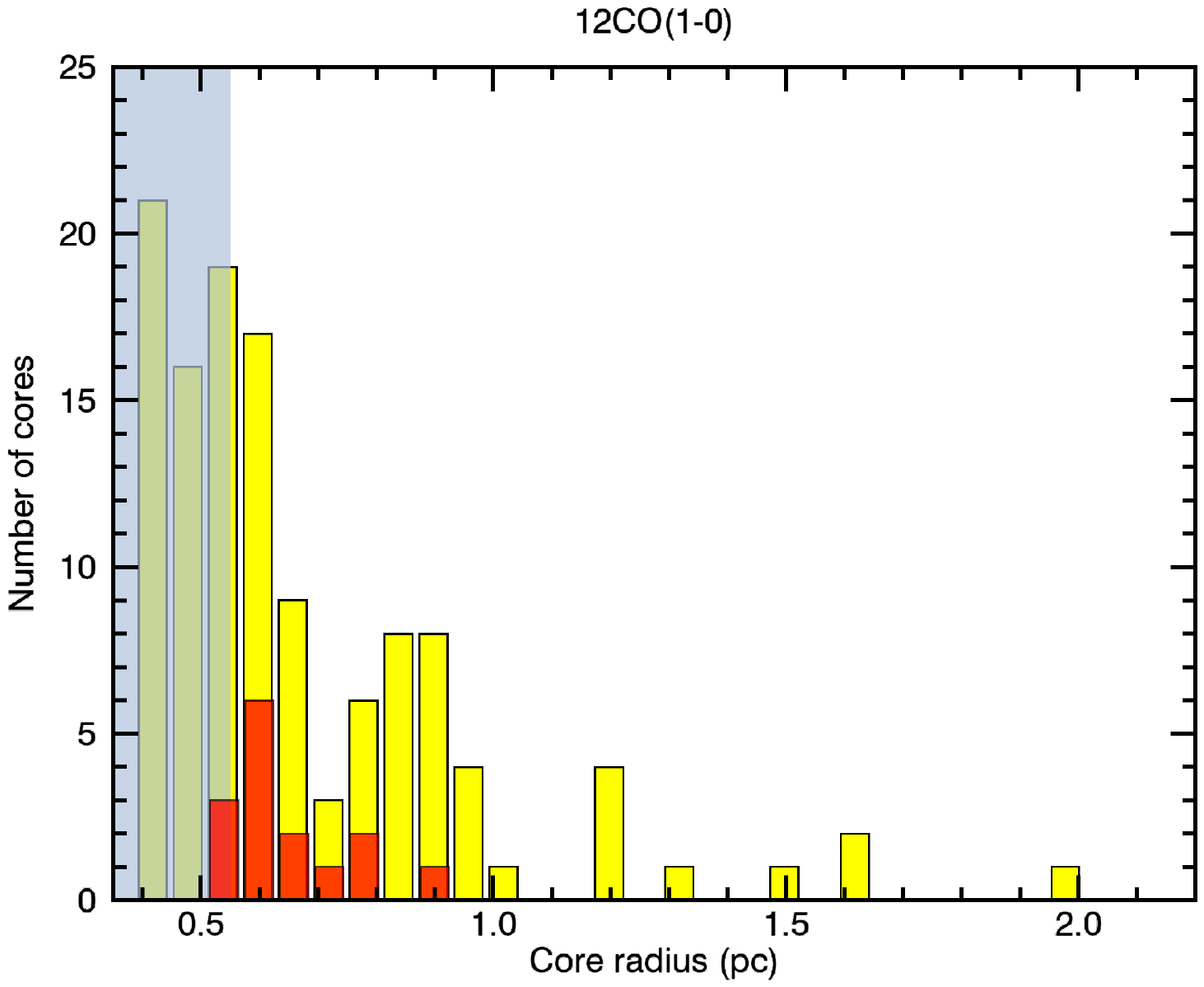}
\includegraphics[scale=0.5]{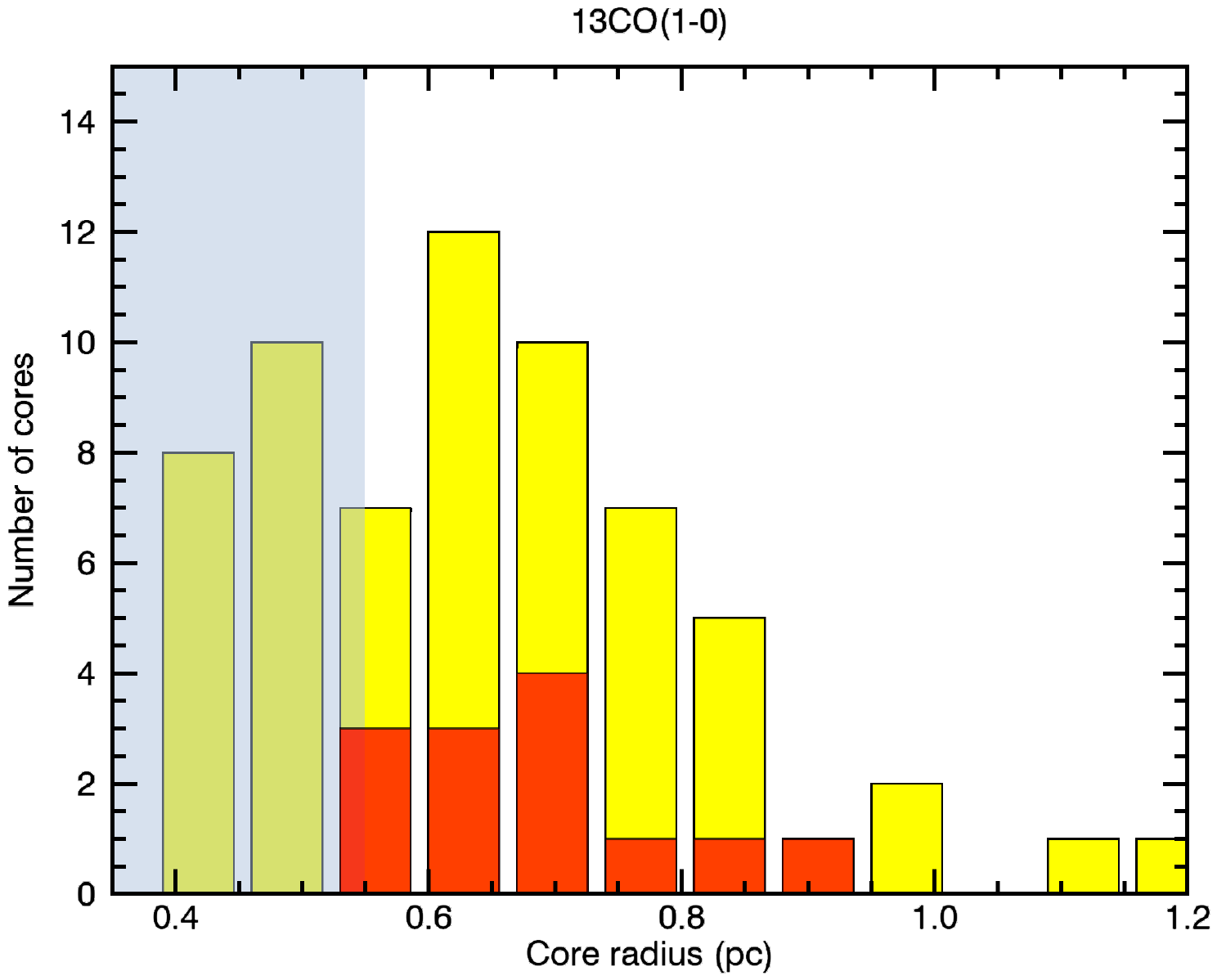}
\centering
\caption{Histograms of $^{12}$CO (left) and $^{13}$CO (right) core radii with (red) and without (yellow) YSOs. The shaded blue indicates the region where the observational resolution affects the sizes.}
\label{12COradius_dis}
\end{figure}

\begin{figure*}
\centering
\includegraphics[scale=0.5]{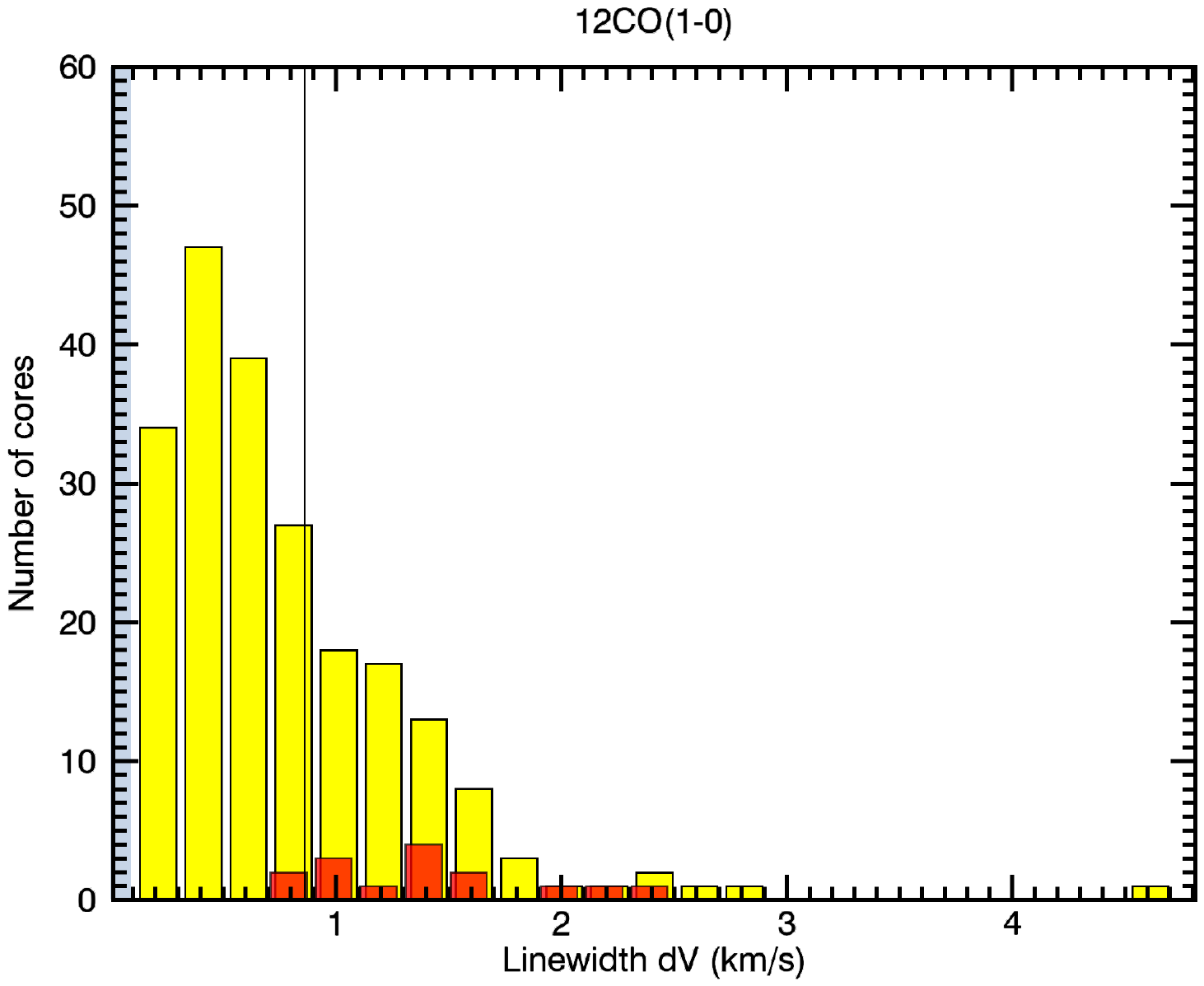}
\includegraphics[scale=0.5]{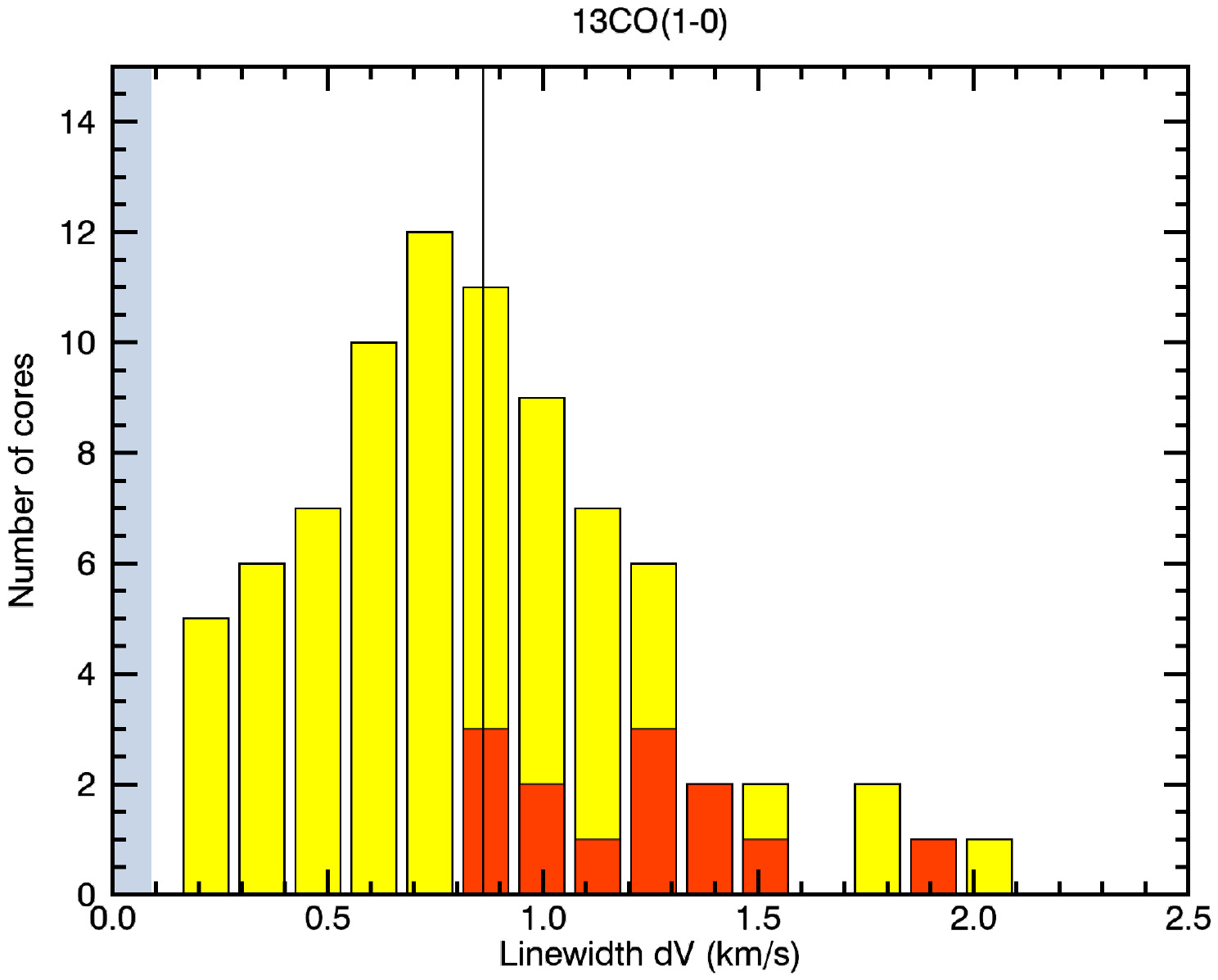}
\centering
\caption{Histograms of $^{12}$CO (left) and $^{13}$CO (right) core linewidths with (red) and without (yellow) YSOs. Thermal and turbulent molecular cores are separated by the vertical lines at 0.861\,km\,s$^{-1}$ ($^{12}$CO) and 0.863\,km\,s$^{-1}$ ($^{13}$CO). The shaded blue indicates the region where the observational resolution affects the linewidths.}
\label{13COlinewidth_dis}
\end{figure*}

The size and linewidth determinations can be biased by the limited instrumental
resolution and sensitivity. This can be particularly important when the extent
of the intensity distribution is comparable to the instrumental profile, e.g.,
the structures have sizes similar to or less than the beam width. \citet{rosolowsky06} applied a correction for this bias by extrapolating the size,
linewidth, and flux to a zero noise level and then deconvolving the instrumental
resolution. To determine the deconvolved sizes, for instance, they subtracted
the rms beam size from the extrapolated spatial moments along the major and
minor axes of the clump, ($\sigma_x$, $\sigma_y$), in quadrature. They found that
without deconvolution, the resolution effect can exaggerate the clump size by
$\sim$40$\%$, and that applying their deconvolution method recovers the size to
within 10$\%$ for S/N ratios greater than 5.  The chief drawback to this approach
is the required extrapolation to the zero intensity level, which is much less
reliable for blended structures (such as clumps within a molecular cloud) than
for isolated, discrete structures. For instance, \citet{wong08} showed that
extrapolation increases the $^{13}$CO clump masses in the RCW 106 GMC by roughly
an order of magnitude, violating the limit imposed by the total map flux. On
the other hand, deconvolution without extrapolation would lead to an underestimation of the
sizes of structures and lead to many of the smallest structures (what we refer to as cores) having unreliable size because their apparent sizes (above a noise threshold) are less than the beam size. Thus,
instead of attempting to correct for instrumental resolution, we indicate the
regions (as shaded blue) where the resolution significantly affects the values in the histogram and correlation
plots (Figures \ref{12COradius_dis}, \ref{13COlinewidth_dis} and \ref{r-v}). The channel width divided by $\sqrt{8\,{\rm ln}\,2}$ is taken as resolution in $\sigma_v$, and the beam width multiplied by 1.91 and divided by $\sqrt{8\,{\rm ln}\,2}$ is taken as resolution in radius \citet{wong17}. In our analysis the sizes of all identified dendrogram trunks are above the resolution limit ($\sim 0.55$\,pc), hence these structures are large enough to resolve the clumps. About 30$\%$ of the dendrogram leaves are having a size below the resolution limit which are the smallest structures in the dendrogram. We do not consider those as reliable structures.

\begin{figure}
\centering
\includegraphics[scale=0.3]{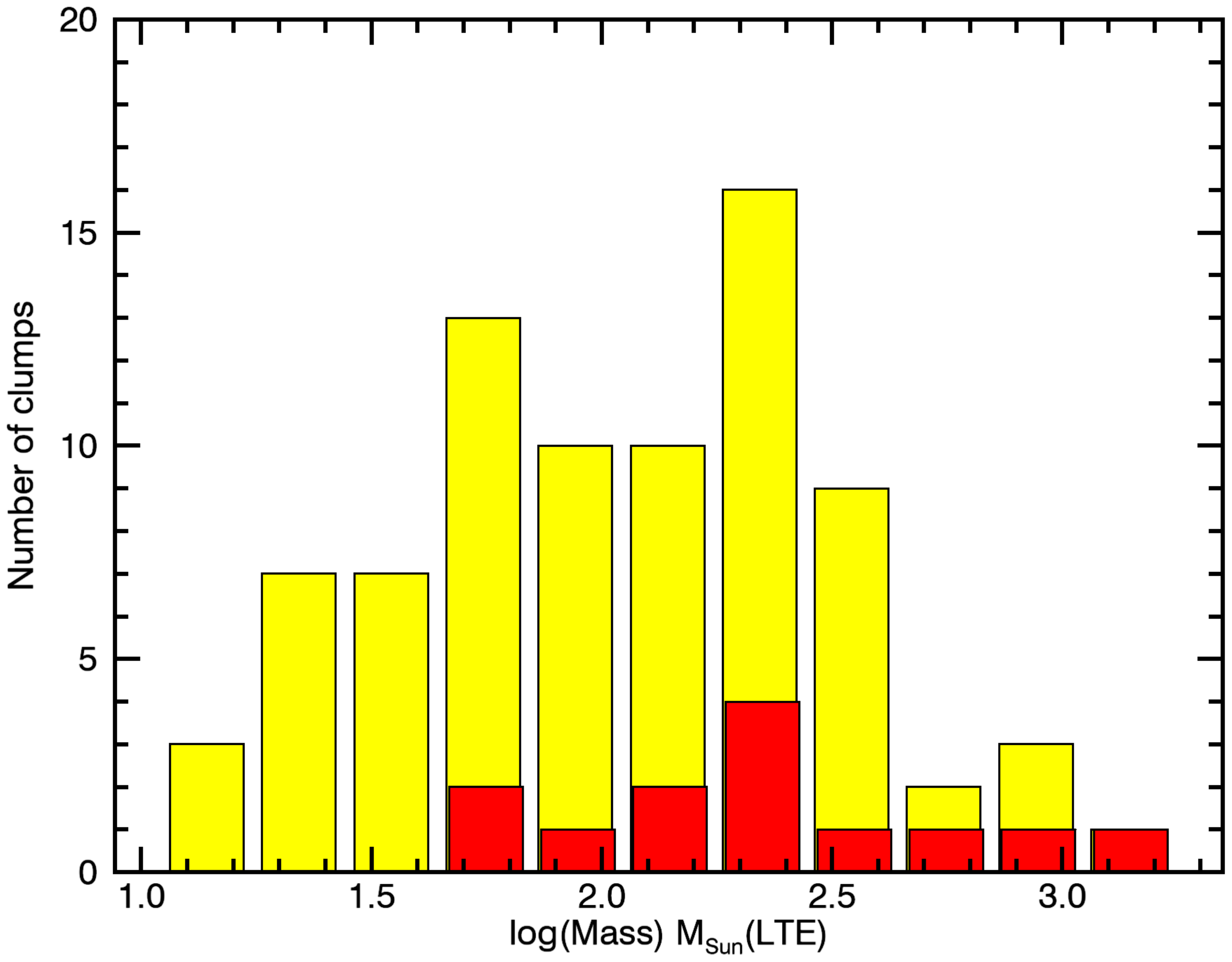}
\centering
\caption{Histograms of $^{13}$CO core LTE masses with (red) and without (yellow) YSOs.}
\label{13COmass}
\end{figure}
\section{Results}
\subsection{Molecular clump properties and association with YSOs}

Our dendrogram analysis on $^{13}$CO(1-0) and $^{12}$CO(1-0) emission
provides the cloud properties such as size, velocity dispersion, and
mass of N\,55. We use these properties to understand the close
relation of star formation with molecular cores. We derive the
apparent velocity width $\Delta V$, FWHM, of each clump by multiplying the
velocity dispersion $\sigma_v$ with $2\sqrt{2{\rm ln}2}$, assuming a
Gaussian distribution.

We compare the sizes and velocity widths of molecular clumps in N\,55 with YSOs and
those without YSOs. For this purpose, we choose the molecular cores in
$^{13}$CO(1-0) and $^{12}$CO(1-0) emission that are identified as
leaves (Tables \ref{t_lines4} and \ref{t_lines3}) with {\it astrodendro} where the stars are assumed to be formed. The size distribution of
$^{13}$CO(1-0) and $^{12}$CO(1-0) cores are given as histograms in
Figure \ref{12COradius_dis}. Radii of $^{13}$CO(1-0) dense cores with YSOs range from
0.55 to 0.9\,pc and those with larger 0.95--1.2\,pc do not show
YSOs. In Figure
\ref{12COradius_dis}, the shaded blue indicates the region where the
observation resolution affects the sizes. About 30$\%$ of CO cores have sizes
less than the instrumental resolution, those we do not consider as
relevant structures in our analysis. The positions of all identified YSOs and associated molecular
clumps are given in Table \ref{YSO_clump}.

The velocity width distributions for $^{12}$CO(1-0) and $^{13}$CO(1-0)
emission are shown in Figure \ref{13COlinewidth_dis}. Those with YSOs
are shown (in red) for comparison. We note that YSOs are embedded
within $^{13}$CO(1-0) dense cores with larger velocity widths
(0.9--2.0\,km\,s$^{-1}$). The distribution of $^{12}$CO(1-0) velocity widths shows
a peak at 0.4\,km\,s$^{-1}$ and the velocity width tail extends up to
4.6\,km\,s$^{-1}$. The $^{12}$CO(1-0) velocity widths for star forming cores
range from 0.8--2.5\,km\,s$^{-1}$. The $^{13}$CO(1-0) velocity widths peak at
0.8\,km\,s$^{-1}$ and show a velocity width tail extending up to 2.1\,km\,s$^{-1}$. The
large velocity width in general suggests the effect of internal
turbulence in molecular cores. If the observed velocity width is
higher than the thermal velocity width, the molecular cores are
expected to be dominated by non-thermal motions.     

According to \citet{Myers91} and \citet{ikeda07}, the massive
star-forming cores with turbulent and thermal motion can be
differentiated on the basis of their critical velocity width,
$dv_{cr}$, at which thermal and nonthermal velocity components are equal.
  The cores are considered to be thermal or nonthermal if the nonthermal velocity component is significantly less than
  or greater than the thermal velocity. Therefore, the cores with observed velocity width
greater than the $dv_{cr}$ is considered as turbulent.
In figure \ref{13COlinewidth_dis} we indicate the critical
velocity widths calculated from equation 8 of \citet{ikeda07} for $^{13}$CO(1-0)
and $^{12}$CO(1-0) with the vertical lines at 0.861 and
0.863\,km\,s$^{-1}$ respectively. We note that all star-forming
$^{13}$CO(1-0) and $^{12}$CO(1-0) cores, have linewidths $\ge dv_{cr}$, indicating some effect of turbulence. In Figure
\ref{13COmass}, we compare the distribution of LTE masses of molecular
cores with (red) and without (yellow) YSOs. We find that 50$\%$ of massive
cores ($\ge$500M$_{\odot}$) are associated with YSOs.

\begin{figure}
\centering
\includegraphics[angle=0, scale=0.6]{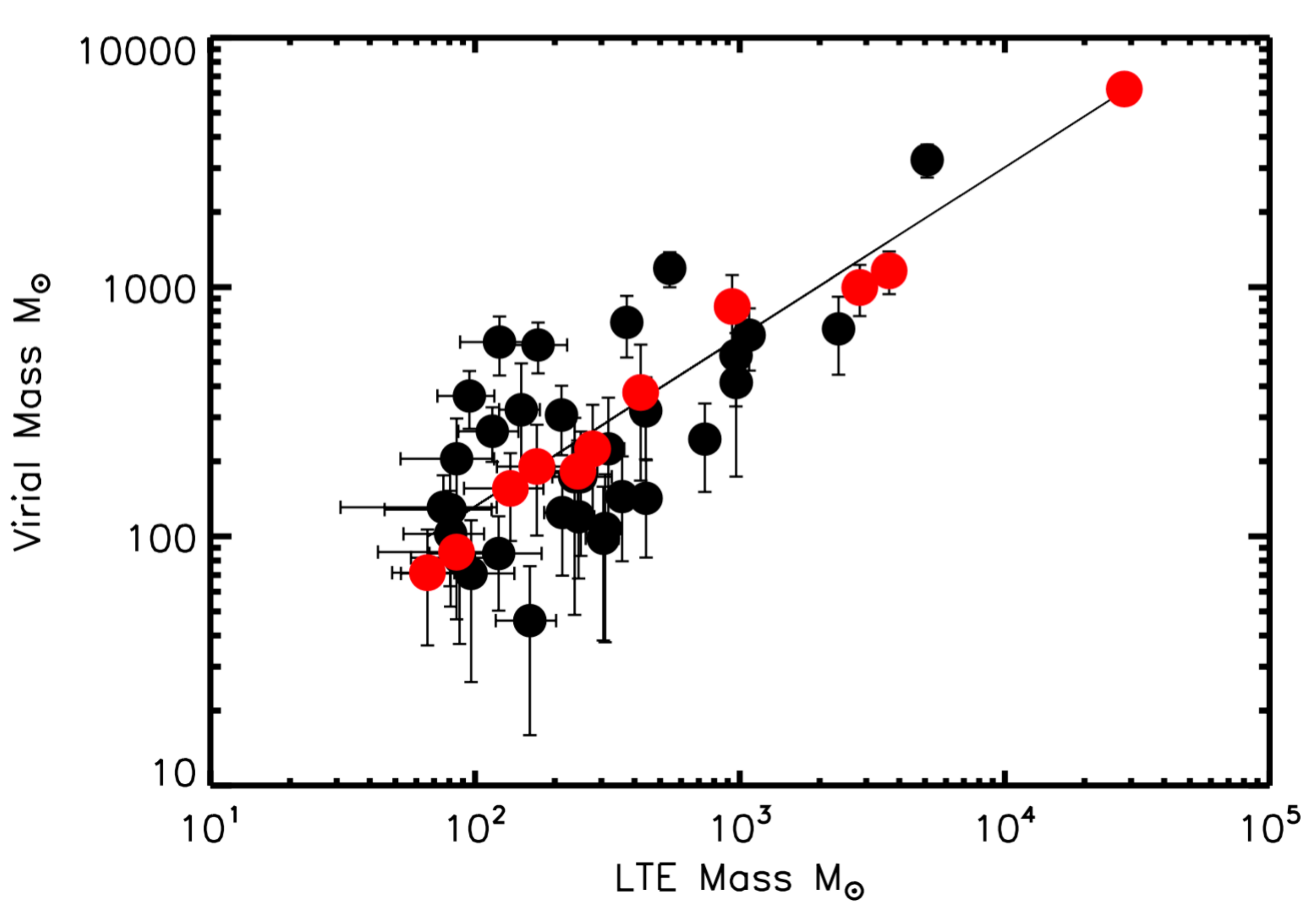}
\centering
\caption{LTE mass versus virial mass relation of $^{13}$CO clumps (black) in N\,55. Those with star formation (with YSOs) are shown in red for comparison.}
\label{virial_ratio}
\end{figure}

Figure
\ref{virial_ratio} shows the relationship between LTE and virial
masses for clumps with YSOs and those without YSOs. The relationship
is fitted with a power law of $M_{{\rm VIR}}$=1.75$M_{{\rm LTE}}^{0.68\pm0.12}$ with a correlation coefficient 0.8 which
indicates a tight correlation. The ratio of LTE mass to virial mass
shows an average value of 1.12, including all star-forming and
non-star-forming clumps. The star-forming clumps cluster around the
power law line $M_{{\rm VIR}}$=1.75$M_{{\rm LTE}}^{0.68\pm0.12}$. Three massive clumps with YSOs have LTE
masses three times larger than their virial masses, and all other
star-forming clumps have LTE masses consistent with virial
masses. These results indicate that most of the star-forming and
non-star-forming cores are gravitationally bound.

\subsection{Scaling relations}

\subsubsection{Size-linewidth relation}

\begin{figure*}
\centering
\includegraphics[scale=0.7]{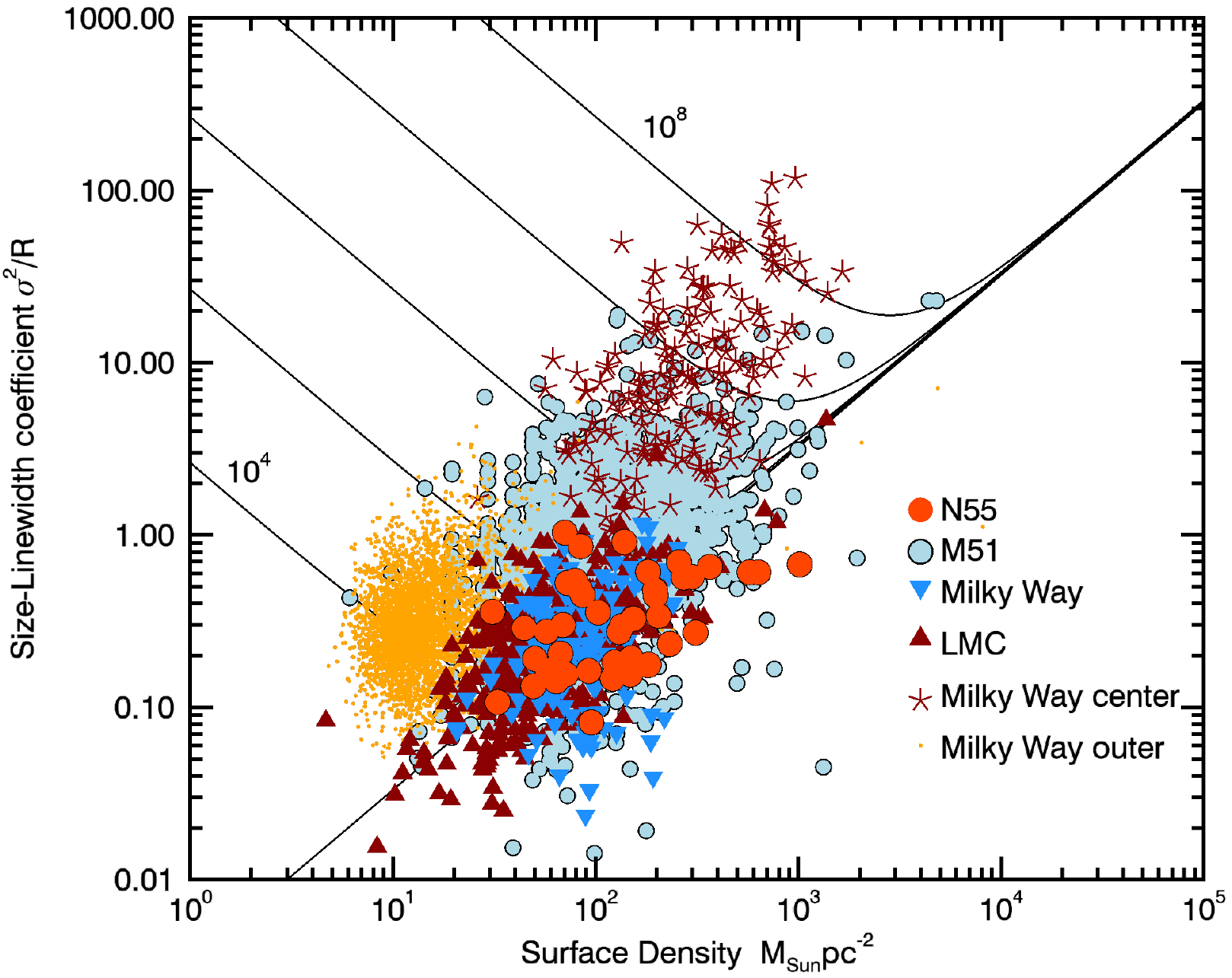}
\centering
\caption{Size-linewidth coefficient $\sigma^2$/R is plotted against mass surface density in M$_{\odot}$\,pc$^{-2}$ for N\,55 $^{13}$CO(1-0) clumps. Milky Way GMCs \citep{heyer09, heyer01, oka01}, M51 \citep{colombo14} and the LMC molecular clouds from Mopra observations by \citet{wong11} are shown for comparison. The $\sigma^2$/R for N\,55 clumps, Milky Way GMCs \citep{heyer09} and the LMC clouds roughly show a linear relation to the mass surface density, indicating that clouds are virialized with negligible external pressure (along the solid diagonal line). The solid curves are pressure bounded virial equilibrium curves for external pressure ranging from $P$/k$_{\rm B}$ = 10$^4$-10$^8$ cm$^{-3}$K. The clouds in the Milky Way centre and Milky Way outer show large velocity dispersion and confined by external pressure.}
\label{size-linewidthcoeff}
\end{figure*}

\begin{figure}
\vspace{0.8 cm}
\centering
\includegraphics[angle=0, scale=0.7]{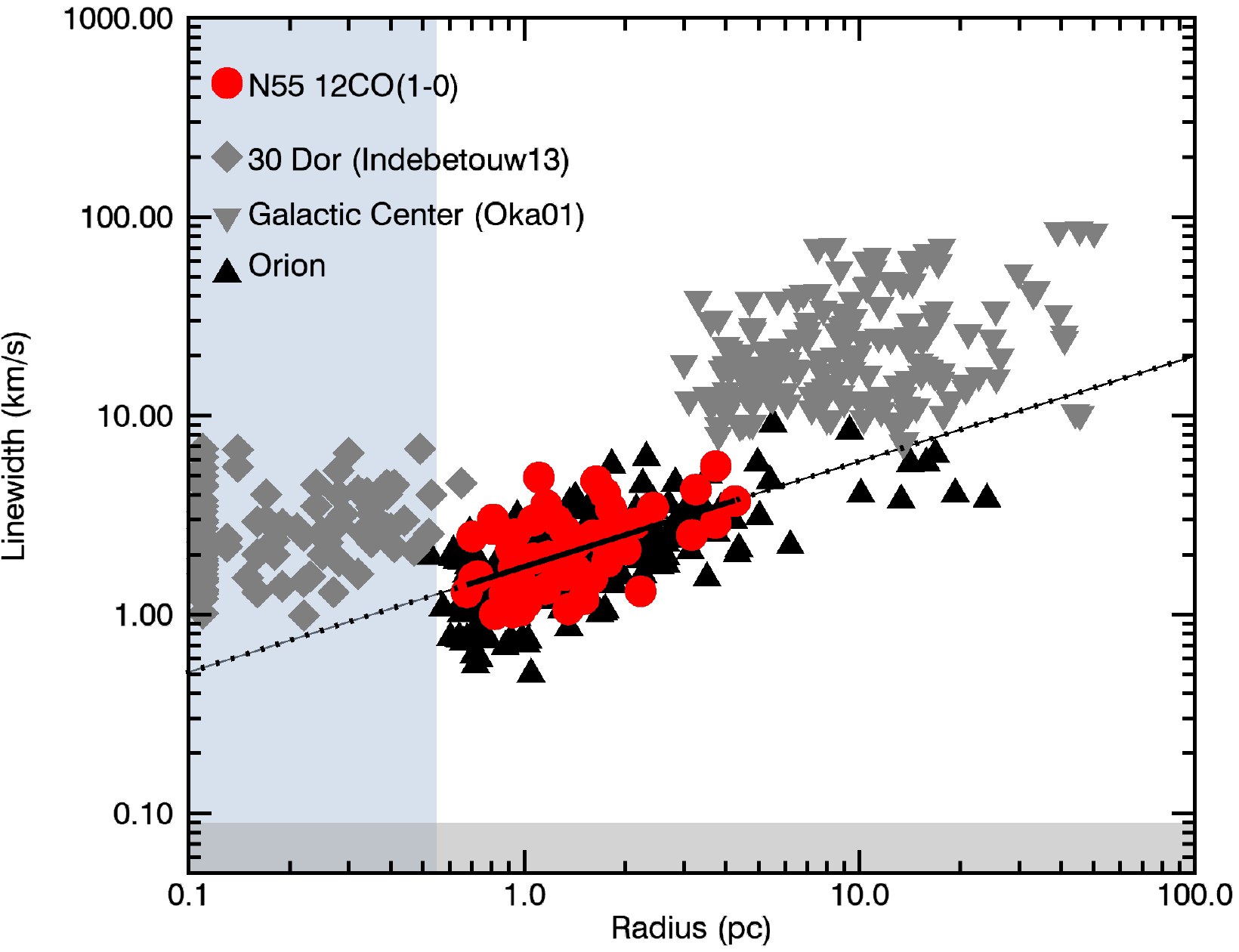}
\centering
\caption{Linewidth versus radius relation for $^{12}$CO(1-0) in N\,55 compared to that of Orion molecular cloud and 30\,Doradus. The N\,55 clumps follow a power law relation, $\Delta V \propto R^{0.5}$, which agrees well with the Orion clumps. The shaded blue indicates the region where the observational resolution affects the sizes.}
\label{r-v}
\end{figure}

In order to check the virial boundedness of the clumps we test
  the relationship between the mass surface density, size and
  linewidth of the clumps \citep{heyer09}, since the virial theorem links those three basic parameters. \citet{Leroy15} have reported a diagnostic diagram
  relating the size-linewidth coefficient $\sigma_{v}^{2}$/$R$ with
  mass surface density to distinguish the pressure-bound clouds from
  gravity-bound. For this relation, the mass surface density is obtained from the LTE assumption.
  In figure \ref{size-linewidthcoeff} we show the
  $\sigma_{v}^{2}$/$R$ - surface density relation for N\,55
  $^{13}$CO(1-0) clumps, where the mass densities are derived from the
  LTE assumption. For comparison we show the Milky Way clouds
  \citep{heyer09} which are observed with spatial scale $3.5-13$\,pc and spectral resolution 1\,km\,s$^{-1}$, the Milky Way center clouds \citep{oka01} observed with 1.4\,pc spatial and 2\,km\,s$^{-1}$ spectral resolutions, the Milky Way outer clouds \citep{heyer01} observed with $0.4-4.5$\,pc spatial and 0.98\,km\,s$^{-1}$ spectral resolutions, M51 clouds \citep{colombo14} observed with 40\,pc spatial and 5\,km\,s$^{-1}$ spectral resolution, and the LMC clouds
  from $^{12}$CO(1-0) Mopra survey \citep{wong11} with 14.5\,pc spatial and 0.5\,km\,s$^{-1}$ spectral resolutions. The solid
  curves are $\sigma_{v}^{2}$/$R$ as a function of surface density for
  a range of external pressures 10$^4$--10$^8$ cm$^{-3}$\,K
  \citep{Field11}. $\sigma_{v}^{2}$/$R$ roughly shows a linear relation
  with mass surface density for N55 clumps, many Milky Way GMCs and the
  LMC clouds, while the Milky Way center and outer clouds show deviation from the linear trend. If the
clouds are in virial equilibrium, i.e. with low external pressure, the
velocity dispersion of these clouds can be determined from their mass
surface density. This relation is shown by a diagonal solid line
in Figure \ref{size-linewidthcoeff}, where the mass surface density
$\sim$ 330$\sigma^2$/$R$. This relation indicates that N\,55 clumps are virialized with negligible external pressure. In Figure \ref{r-v}, we compares the linewidth
(2$\sqrt{2{\rm ln}2}\times$velocity dispersion) versus radius relation
for $^{12}$CO(1-0) clumps (dendrogram trunks) in N\,55 with Orion,
30\,Doradus and Galactic center clouds. For Orion molecular cloud,
$^{12}$CO(1-0) data of \citet{dame01} was decomposed with {\it
  astrodendro} as for N\,55 clumps. The 30\,Doradus size-linewidth
relation is shown for the $^{12}$CO(2-1) transition derived by
\citet{indebetow13}. The radius-linewidth relation for N\,55 is fitted
with a function $\Delta V$\,$\propto\,R^{0.5}$ with a correlation
coefficient of 0.7. This relation agrees well with the Orion molecular
cloud. The 30\,Doradus and Galactic center clouds show larger velocity
dispersions compared to N\,55 and Orion.

\subsubsection{Virial mass-luminosity relation}

\begin{figure}
\centering
\includegraphics[angle=0, scale=0.7]{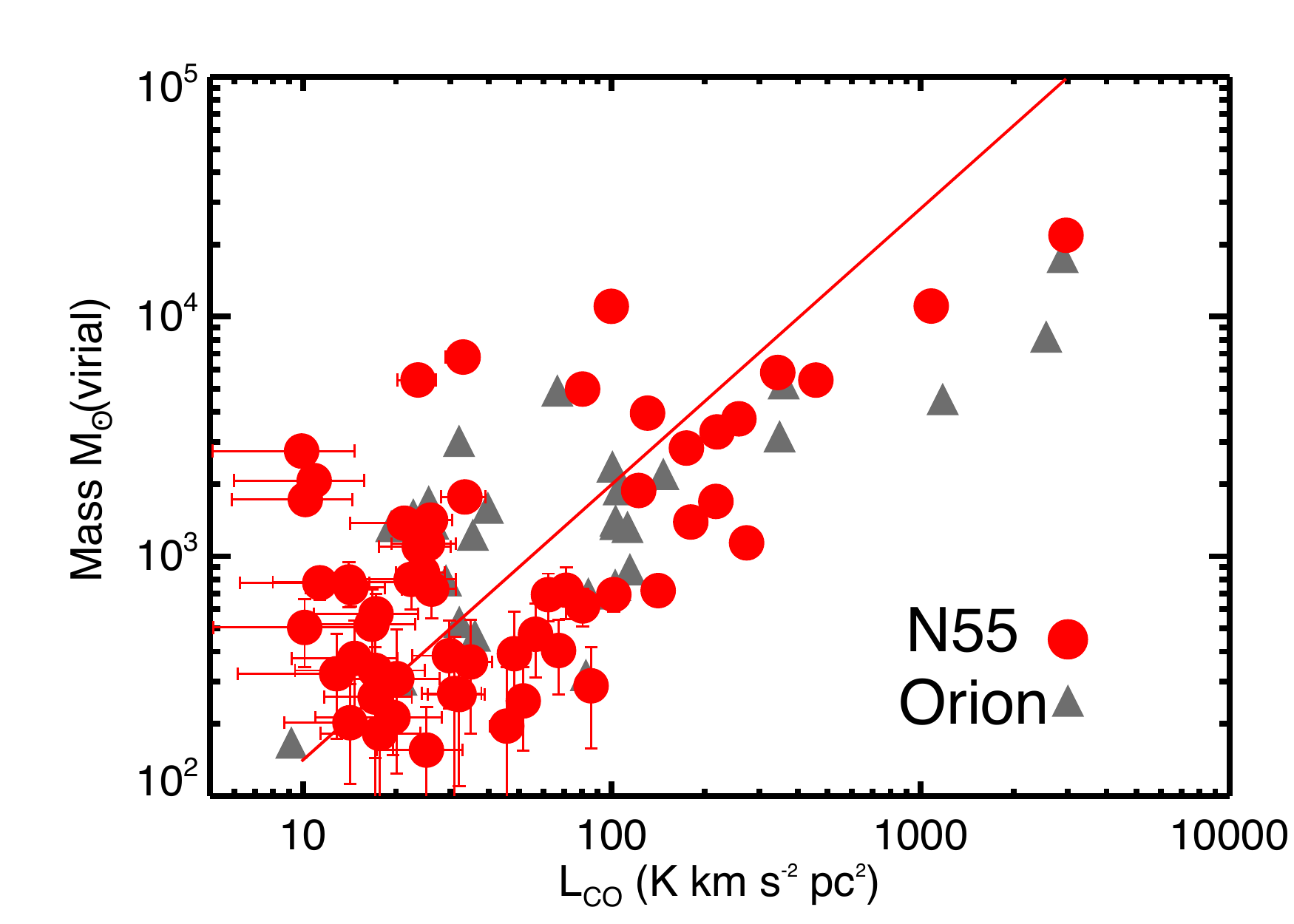}
\centering
\caption{Virial mass versus luminosity relation for $^{12}$CO(1-0) clumps in N\,55 is fitted with a function log($M_{{\rm VIR}}$)=1.0+(1.15$\pm$0.3)\,log($L_{{\rm co}}$) for a correlation coefficient 0.6. For a comparison, the virial mass versus luminosity relation for Orion is shown.}
\label{l-m}
\end{figure}

The Galactic molecular clouds and low-resolution $^{12}$CO(1-0)
observation of the LMC molecular clouds show a power law correlation
between $^{12}$CO luminosity and virial mass. \citet{solomon87} demonstrated how to use the
mass-luminosity relation of optically thick $^{12}$CO(1-0) lines as a calibrator
for total cloud mass, i.e. molecular hydrogen mass. They found a tight
correlation between $^{12}$CO(1-0) luminosity and virial mass with a
power law index of 0.81 for nearly 278 Galactic molecular clouds.
Figure \ref{l-m} compares the virial mass versus the $^{12}$CO(1-0) luminosity relation of N\,55 clumps with the Orion molecular cloud obtained with the NANTEN 4\,m telescope \citep{nishimura15}. The N\,55 clouds show a power law relation between
virial mass and CO luminosity (correlation coefficient 0.6) with a scatter of more than an order of magnitude. The mass-luminosity
relation of N\,55 in Figure \ref{l-m} is fitted with a function
log($M$)=($1.15\pm0.3$)\,log($L_{\rm CO})+1.0$. The Orion cloud shows a power law index of 0.55.

\subsection{Cloud mass spectrum}

\begin{figure*}
\centering
\includegraphics[scale=0.5, angle=0]{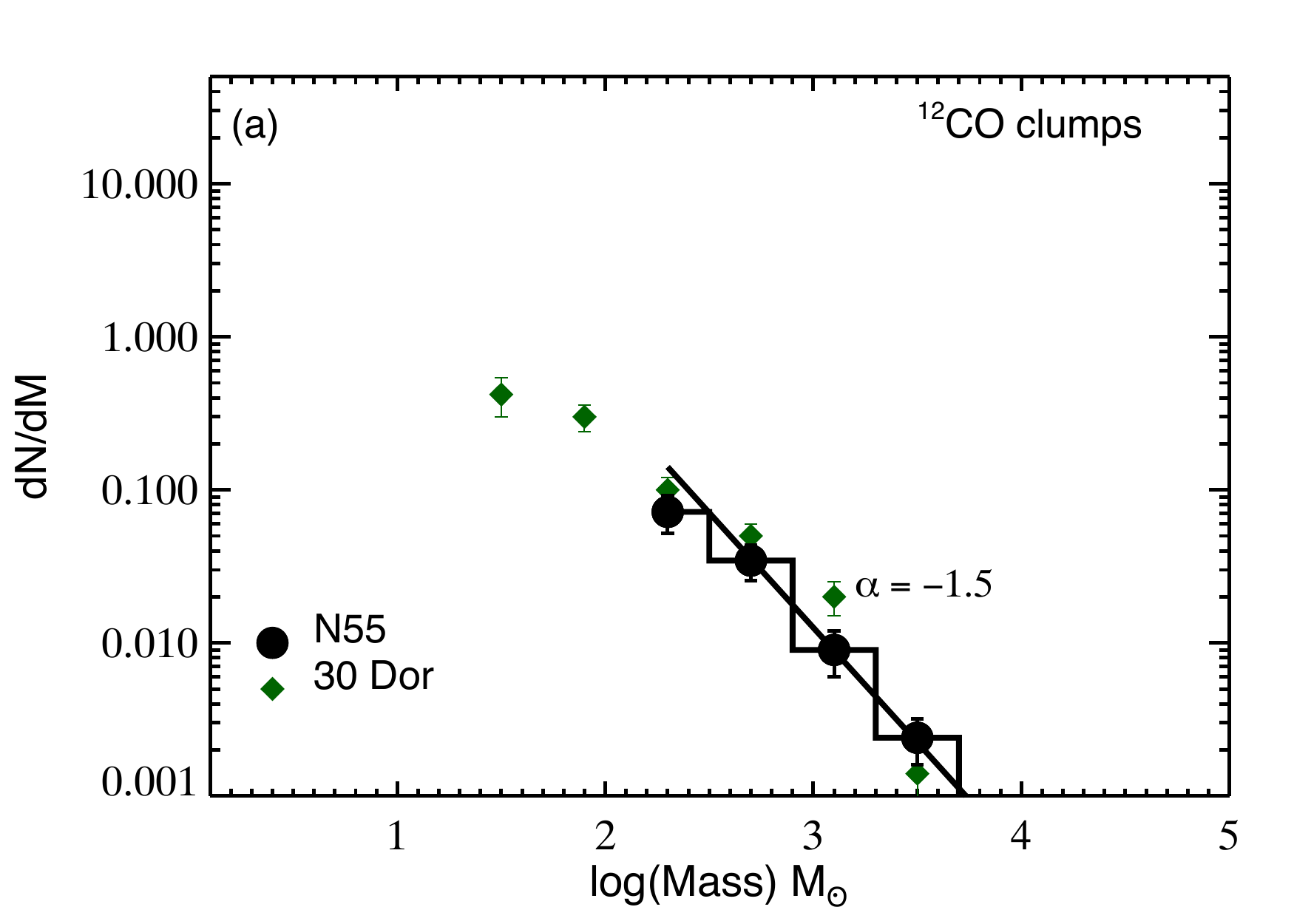}
\includegraphics[scale=0.5, angle=0]{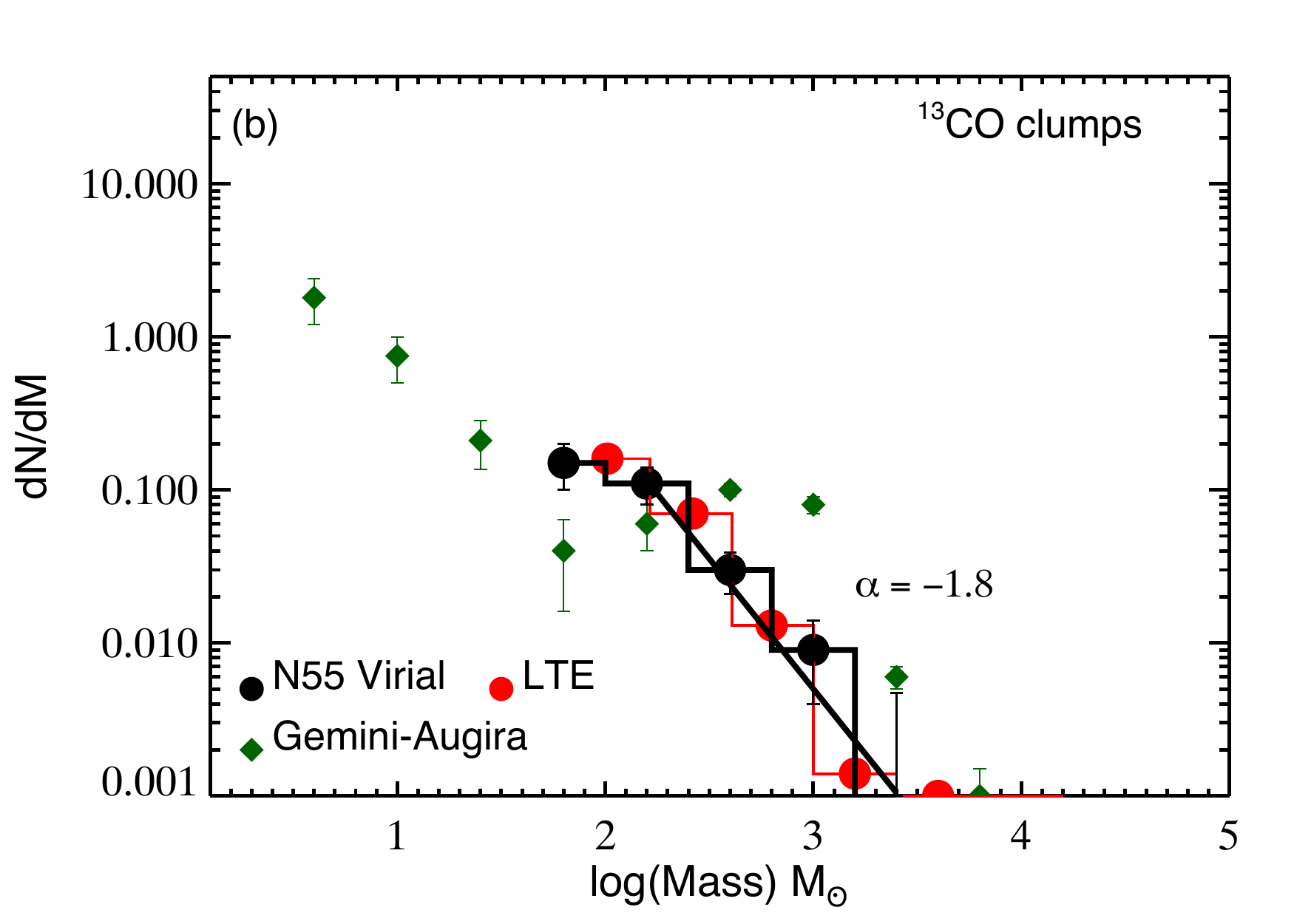}
\centering
\caption{The $^{12}$CO(1-0) and $^{13}$CO(1-0) clump mass spectra obtained from dendrogram trunks are shown in (a) and (b) respectively. The masses are derived from the virial and LTE assumptions. The best fit power-law index for the function $dN/dM \propto M^{-\alpha}$ is $1.5 \pm 0.15$ for $^{12}$CO(1-0) and $1.8 \pm 0.25$ for $^{13}$CO(1-0). For comparison, the clump mass spectra of 30\,Doradus \citep{indebetow13} and Gemini-augira clouds \citep{kawamura98} are given.}
\label{trunks}
\end{figure*}

\begin{figure*}
\centering
\includegraphics[scale=0.5, angle=0]{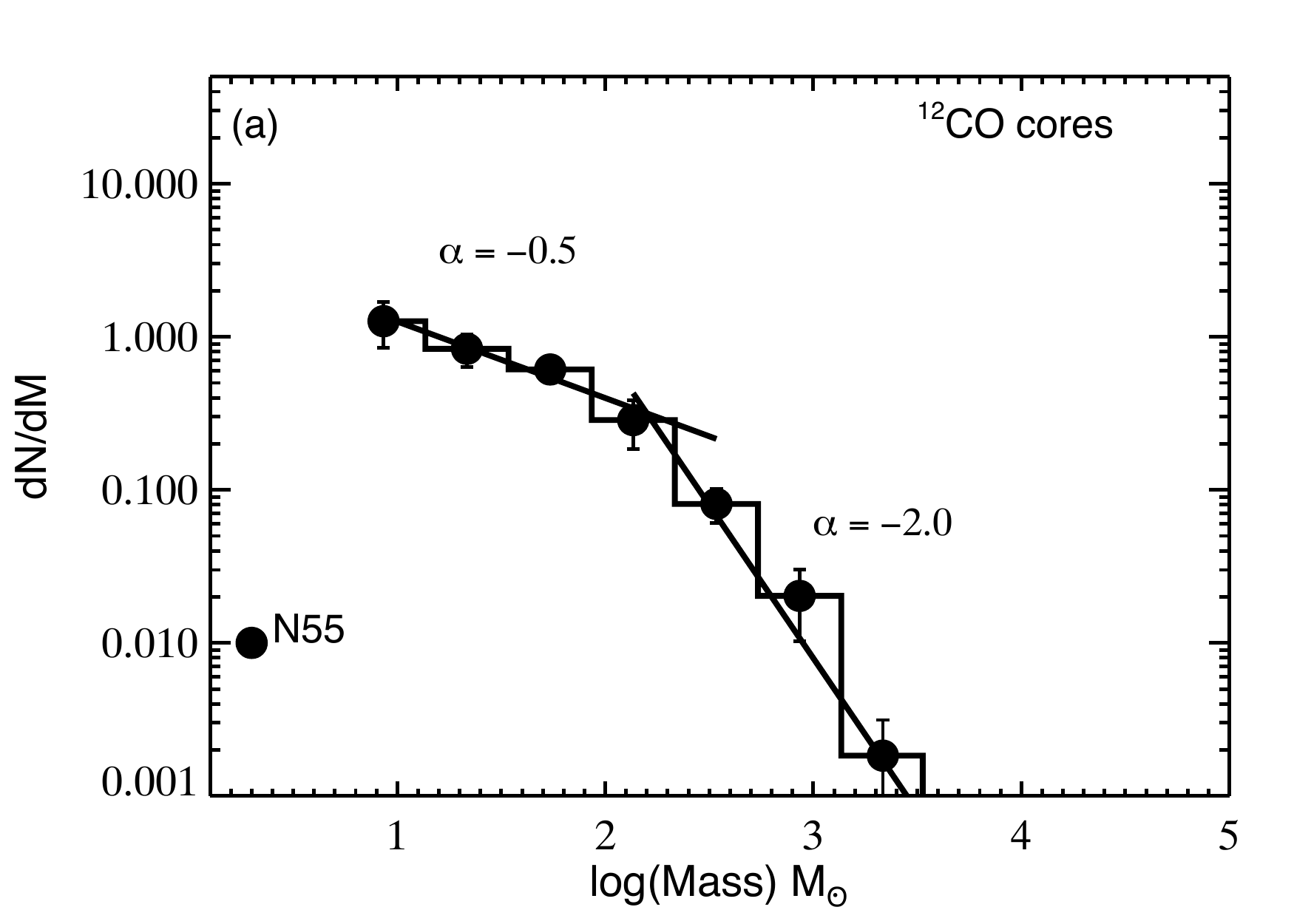}
\includegraphics[scale=0.5, angle=0]{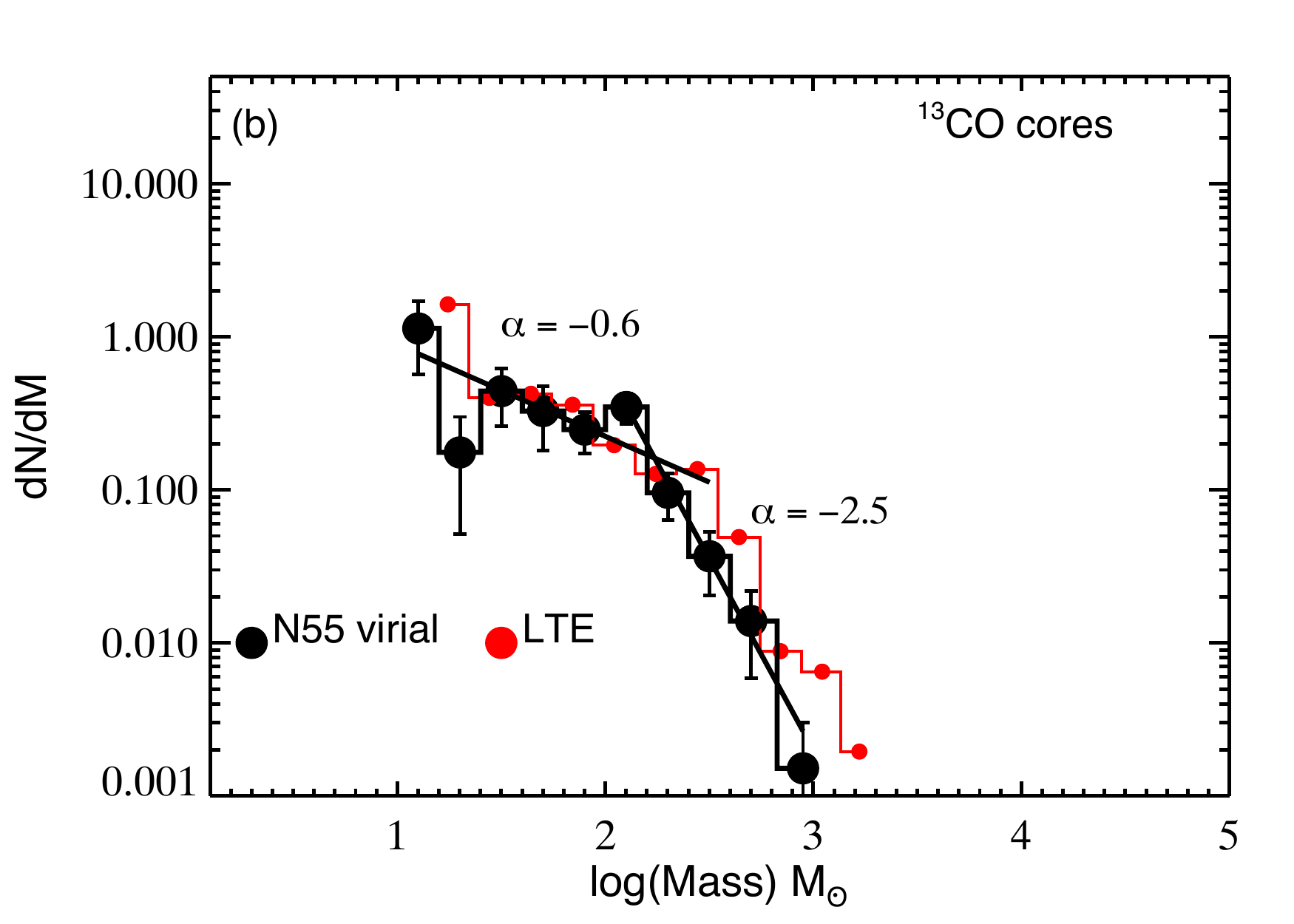}
\centering
\caption{The $^{12}$CO(1-0) and $^{13}$CO(1-0) core mass spectra obtained from dendrogram leaves are given in (a) and (b) respectively. a) The best fit power-law index for $^{12}$CO(1-0) in low-mass end is $\alpha$=$0.5 \pm 0.05$, and in high-mass end $\alpha$=$2.0 \pm 0.3$. b) The best fit power-law index for the $^{13}$CO(1-0) core mass spectrum is $0.6 \pm 0.2$ in low-mass end, and $2.5 \pm 0.4$ in high-mass end.}
\label{leaves}
\end{figure*}

The mass spectrum of a distribution of molecular clumps in a cloud is
usually represented in a differential form of cloud numbers N$_i$ in the
mass range between $M_i$ and $M_i + dM$, where $dM$ is mass bin. This
can be represented as $dN/dM = f(M)$. It is found that the molecular cloud mass spectrum is often well
described by a power law:
\begin{equation}
\frac{dN}{dM} \propto M^{-\alpha} 
\label{Masseqn}
\end{equation}

The above differential form of the power law can be sensitive to the
adopted mass bins, hence a cumulative or a truncated power law is
often adopted to represent the mass distribution, $dN/d{\rm ln}M \propto
(M_m/M)^{\alpha}$ \citep{kawamura98, indebetow13}. Here $dN$ is the
number of clouds in the mass range $M_i$ and $M_i + d{\rm ln}M$, and $M_m$
is the upper mass. However, it is much easier to recognize the
behavior of a mass spectrum in differential form, as in equation
\ref{Masseqn}. Therefore, in our analysis we use the differential form
of clump mass spectrum i.e. $dN/dM = f(M)$. We find that index of
the power law is not affected by the mass bins we adopted.

From the dendrogram analysis of $^{12}$CO(1-0) and $^{13}$CO(1-0) maps, we
have obtained two sets of structures, molecular cores ({\it leaves}) and clumps ({\it trunks}). We use these leaves and trunks separately to construct the
mass spectra and compare their trends in Figures \ref{trunks} and
\ref{leaves}. We compare the mass spectra of N\,55 from the virial
as well as LTE masses calculated for $^{12}$CO(1-0) and $^{13}$CO(1-0). The trunks
reveal the high mass end of mass spectra in both $^{12}$CO(1-0) and
$^{13}$CO(1-0) which are fitted with the functions $dN/dM \propto
M^{-1.5\pm0.15}$ and $dN/dM \propto M^{-1.8\pm0.25}$ respectively. In
Figure \ref{trunks}\,a, for comparison we show the mass spectrum of
the star-forming region 30\,Doradus in the LMC from $^{12}$CO(2-1)
clumps observed with ALMA \citep{indebetow13} along with the
$^{12}$CO(1-0) clump mass spectrum in N\,55. In
Figure \ref{trunks}\,b, we present the clump mass spectrum for
$^{13}$CO(1-0), and compare the power law behavior with Gemini-Augira
cloud (in the same mass range 10-10$^4$M\,$_{\odot}$) in the Milky Way
\citep{kawamura98}. The dendrogram leaves are tracers of the low-mass
cores along with a few high-mass cores which clearly show a turnover
below 200\,M$_{\odot}$ (Figure \ref{leaves}) in the mass spectrum. The
low-mass end of $^{12}$CO(1-0) mass spectrum, $\le$ 200\,M$_{\odot}$, is
fitted with a power law index of 0.5$\pm$0.1 and that for $^{13}$CO(1-0)
is fitted with a power law index of 0.6$\pm$0.2. The steep high-mass part
shows power law indices 2.0$\pm$0.3 and 2.5$\pm$0.4 for $^{12}$CO(1-0)
and $^{13}$CO(1-0) cores respectively.

\section{Discussion}
\subsection{Clump physical properties: size-linewidth relation and mass spectrum}
We quantify the molecular clump properties at smaller scales
(sub-parsec $\sim$0.1\,pc) and probe the GMC characteristics' close
relationship to star formation in N\,55 using ALMA observations. In
order to investigate, how the N\,55 GMC properties differ from those of the Galaxy, we compare the
size-linewidth, virial mass-luminosity and clump mass function
relations of clouds in N\,55 with the Orion molecular cloud, Milky
Way outer and inner clouds and Gemini-Augira regions (Figures
\ref{r-v} - \ref{leaves}). As we have already noted in section 1, the Galactic
molecular clouds and many extragalactic clouds which are observed at
resolutions of $>$10\,pc show a power law relation between clump size
and velocity dispersion $\sigma_v$: $\sigma_v \propto R^{\beta}$. 
Does this size-linewidth power law relation hold or breakdown in
sub-parsec scales for the LMC? Do we find any deviation in
size-linewidth power law behavior at lower metallicity, where the
clumps can be highly dissociated by hard UV radiation or perturbed by
shocks due to massive star formation?

\citet{indebetow13} and \citet{nayak16} have reported cloud properties of 30\,Doradus using
$^{12}$CO(2-1) observations with ALMA. They noticed a large
scatter in the size-linewidth relation with relatively high velocity
dispersion in 30\,Doradus compared to the Galactic clouds, indicating
the effect of external pressure in addition to gravitational
equilibrium. This behavior can be checked in other star-forming
regions in the LMC. The size-linewidth relation in Figure \ref{r-v}
indicates that the N\,55 molecular clumps show very similar power law
relationship as in the Orion cloud. The Galactic center and
30\,Doradus clouds show large velocity dispersions which the authors
say may be due to the effect of external pressure. In Figure
\ref{size-linewidthcoeff}, we try to evaluate whether any effect of
external force is required for equilibrium confinement of N\,55
molecular clumps in addition to gravity. Like many Milky Way GMCs and
the LMC molecular clouds from the Mopra survey, the N\,55 clouds
reported in this work cluster around the equilibrium
line (solid diagonal line in Figure \ref{size-linewidthcoeff}) as
defined by \citet{Leroy15}. This relation suggests that N\,55 clumps
are gravitationally bound. Clouds clustering above this line include
GMCs in the outer and center regions of the Milky Way, which require
high external pressure for confinement in a high turbulent medium. We
note that in N\,55 clouds the average ratio of virial mass to LTE mass
is 1.12, where the LTE mass versus virial mass relation is fitted with
a power law index 0.68$\pm$0.12 which is consistent with the power
index found for Orion cloud (i.e. gravitationally bound)
\citep{ikeda09b, liu12}.

The core mass distribution is another important characteristic of the
molecular cloud population which has a significant impact on
star-forming clouds, because of its similarity with the stellar
initial mass function. For the inner Milky Way clouds, the mass
distribution follows a power law relation $dN/dM \propto M^{-\alpha}$
where $\alpha=1.5$. The surveys of the Local Group galaxies have
revealed the power law behavior of cloud mass spectrum \citep{wong11}.
The clump mass spectrum of N\,55 tends to follow a similar behavior as
in 30\,Doradus of the LMC, Gemini-Augira cloud, and many other
star-forming clouds in the Milky Way. We note that ALMA observation of
$^{12}$CO(1-0) and $^{13}$CO(1-0) emission in N\,55 reveal more
massive clumps than in 30 Doradus by $^{12}$CO(2-1) emission. This may
be due to the hard radiation field and preferential destruction of
molecular clumps in 30 Doradus.  \citet{kawamura98} reported a power
law index of 1.4--1.9 for Gemini-Augira clouds. Similar studies have
been done for Cepheus and Cassiopeia regions with the $^{13}$CO(1-0)
survey by \citet{Yonekura97}. They have compared the mass spectrum
behavior of Cepheus and Cassiopeia with several other $^{13}$CO
Galactic clouds and reported a power law index in the range of
1.5--1.8 for most of the $^{13}$CO clouds with the mass range
10--10$^5$M$_{\odot}$. This power law behavior of the clump mass
spectrum is consistent with many star-forming clouds in the Milky Way,
such as Orion \citep{ikeda09a, johnstone00, johnstone06},
$\rho$Ophiuchi cloud \citep{motte98}, and Taurus
\citep{onishi02}. These studies have shown the power law index of
2.0--3.0 in the high-mass part. Our studies show that the mass
spectrum index is not dependent on the environments sampled. This
study confirms a universal behavior of the clump mass function at
smaller spatial scales in a sub-solar metallicity galaxy.

\subsection{X$_{{\rm CO}}$ factor}

We determine the CO-to-H$_2$ conversion factor for N\,55 molecular
clouds using our $^{12}$CO(1-0) observations.  Since the determination
of molecular hydrogen mass is fundamental to understand the physics of
star formation, and a direct detection of bulk H$_2$ mass is almost
impossible, the CO luminosity-mass relation is widely used as a
calibrator for measuring the CO-to-H$_2$ mass conversion factor, X$_{\rm CO}$ 
\citep{solomon87}. The reason for using the CO luminosity-mass relation as a calibrator for X$_{\rm CO}$ is that, for gravitationally bound clouds the ratio of virial mass to
$^{12}$CO luminosity is directly proportional to the
square root of H$_2$ density and inversely proportional to the average
brightness temperature \citep{solomon91}. In our analysis, the masses of N\,55 clumps are determined by
LTE and virial assumption of $^{12}$CO and $^{13}$CO emission. The
consistency between LTE mass and virial mass of the clumps along with
a virial ratio of 1.12, and the virial equilibrium relationship in the
$\sigma^2$/R versus mass surface density plot (where the mass surface
density is determined using the LTE method) strongly suggest that N\,55
clumps are gravitationally bound. \citet{fukui08} reported a power law relation
between $^{12}$CO(1-0) luminosity and virial mass with an index of
1.2$\pm$0.3 for the LMC molecular clouds. This relation yields an
X$_{\rm CO}$ factor of 7$\times$10$^{20}$\,cm$^{-2}$(K\,km
s$^{-1}$)$^{-1}$ for a total cloud mass of 28$\times$10$^5$
M$_{\odot}$ and luminosity 2$\times$ 10$^5$ K km\,s$^{-1}$\,pc$^{2}$
using the NANTEN CO survey. Note that the NANTEN CO survey of
  the LMC has a spatial resolution of 2.6$^{\prime}\sim$\,37\,pc. Using the Mopra
  CO data of the LMC at spatial resolution 45$^{\prime\prime}\sim$\,11\,pc, the value of X$_{\rm CO}$ is reported to
  be 4$\times$10$^{20}$\,cm$^{-2}$(K\,km s$^{-1}$)$^{-1}$
  \citep{hughes10}. The virial mass-luminosity relation of N\,55 is
fitted with a power law index of 1.15$\pm$0.3 (Figure \ref{l-m}). This
relation yields an X$_{{\rm CO}}$ factor
6.5$\times$10$^{20}$\,cm$^{-2}$(K\,km s$^{-1}$)$^{-1}$ for a total
cloud mass of 1.2$\times$ 10$^5$\,M$_{\odot}$ and luminosity of
8$\times$ 10$^3$K\,km s$^{-1}$\,pc$^{2}$. This X$_{{\rm CO}}$ factor
value is consistent with the value given in \citet{fukui08}. In Figure
\ref{l-m}, we compare the mass-luminosity relation of the N\,55 clumps
with that of Orion cloud. The mass-luminosity relation of the Orion
cloud yields an X$_{{\rm CO}}$ factor comparable to that found for
Galactic clouds \citep{polk88}.

\subsection{Star formation}

 In order to investigate the efficiency with which
   the stars form within the dense cores, we adopt the YSOs
   identified from {\it Spitzer} and {\it Herschel} observations. In
   section 7.1 we compare the physical properties of the star-forming CO
   cores with the non-star-forming CO cores. We compare the histograms of radius, velocity width, and
   mass of $^{13}$CO and $^{12}$CO cores. The star-forming CO cores tend to have larger masses compared with the
   non-star-forming cores. The velocity widths of the $^{13}$CO and
   $^{12}$CO cores also show a similar trend, where the star-forming
   cores have a velocity width larger than the critical velocity width.
   This may indicate some effect of turbulence.  This result is
   consistent with the study of $^{13}$CO clumps in 30\,Doradus, where
   the massive star formation occurs in clumps with high masses and
   linewidths \citep{nayak16}. The size-linewidth relation of N\,55
 indicates a negligible effect of external pressure, hence the larger
 velocity width of star-forming clouds might be due to high radiation
 fields or shocks caused by nearby massive stars. These findings are
 consistent with massive star forming regions in Galactic clouds such
 as Orion molecular cloud \citep{ikeda09a} and Gemini-Augira cloud
 \citep{kawamura98}.

\citet{gruendl09} and \citet{seale14} identified about 16 young stellar objects
in N\,55 using infrared colors. Among them, 15 are in the field of our
ALMA observation. From \citet{gruendl09} we obtained {\it Spitzer}
photometric magnitudes of seven YSOs. However, we do not have enough
parametric information, such as mass and luminosities for a detailed
study of star formation efficiency. \citet{gruendl09} suggested a
selection criterion 8.0${\,\rm \mu m}$\,$\le$\,[8.0] for massive YSOs in the LMC. This
can be a reasonable approximation for estimating the masses, even
though a strict mass-luminosity relation is not valid. The infrared spectral energy distribution fitting of the YSOs in N\,44 and N\,159 of the LMC have shown that, the brightest YSOs with 8.0${\,\rm \mu m}$
magnitude $\le$\,[8.0] have masses $\ge$ 8.0M$_{\odot}$ \citep{chen09, chen10}. The brightest
source identified with {\it Spitzer} in our field of observation has a
8.0${\,\rm \mu m}$ magnitude of [7.27$\pm$0.07] which can have a mass
of 20M$_{\odot}$ and a luminosity of 6.0$\times$10$^4$ L$_{\odot}$
according to \citet{chen09} (Table 7). Three sources have 8.0${\,\rm \mu m}$ magnitudes $\sim$\,[8.76, 8.94, 8.24] which represent 
masses in the range 8--15\,M$_{\odot}$, and two others show 8.0${\,\rm \mu m}$
magnitude $\sim$\,[10.17, 10.49] which may have masses below
8.0M$_{\odot}$. These studies indicate that N\,55 is a site for high and
intermediate mass star formation in the LMC.

\begin{table}
\centering
\caption{Position of identified YSOs and associated clumps}
\label{YSO_clump}
\begin{tabular}{@{}ccccccc}
\hline
&  Number   & \multicolumn{2}{c}{Identified YSOs} &   \multicolumn{2}{c}{Associated Clumps} & $^{\rm c}$Clump ID \\
&       &      R.A. (deg)       &    Decl. (deg)       &    R.A. (deg)        &    Decl. (deg) &    \\
\hline
&   1$^{\rm a}$   &  083.0245  &-66.4252    &   83.0270   &  -66.4258  & 23 \\            %   \\
&   2   &  083.0936 & -66.4601    &   83.0936   &  -66.4602 &   53  \\
&   3   &  083.0946 & -66.4523    &   83.0949   &  -66.4523 &  33  \\               %\\
&   4   &  083.1076 & -66.4860    &   83.1077   &  -66.4858 &  37 \\
&   5   &  083.1335 & -66.4542    &   83.1352  &   -66.4547 &  77  \\
&   6   &  083.1371 & -66.4526    &   83.1372  &   -66.4524 &  38   \\             %\\     
&       &            &            &   83.1374  &  -66.4563  &  2   \\
&       &            &            &   83.1404  &  -66.4550  &  55 \\
&   7    &  083.0357 & -66.3870   &   83.0342  &   -66.3869 &   98*   \\%12CO  
&   8$^{\rm b}$   &   082.9186&  -66.4311   &   82.9227  &  -66.4325 & 4    \\
&   9    &  083.0024 & -66.4206   &    82.9999 &   -66.4202  &  72   \\
&        &            &           &    83.0045 &  -66.4207 &   66  \\
&   10    & 82.9945 & -66.4267   &    82.9923   &-66.4275  &   91*  \\%12CO
&   11   &  82.9977 & -66.4287   &    82.9923  &-66.4275 &    \\%12CO
&   12   &  83.0428 & -66.4134   &   83.0435   &  -66.4139 & 28   \\
&   13   &  83.0462 & -66.4151   &   83.0486   &  -66.4152 & 5   \\
&   14   &  82.9979 & -66.4635   &   83.0581 & -66.4684  &   125*  \\%12CO
&   15   &  83.1578 & -66.4699   &  83.1581    &  -66.4703 &  10  \\%12CO

\hline 

\end{tabular}
\parbox{100mm}{
a: {\it Spitzer} YSOs from \citet{gruendl09}.
b: {\it Herschel} YSOs from \citet{seale14}.
c: $^{13}$CO clump number IDs from Table \ref{t_lines4}; those with * are identified in $^{12}$CO emission (Table \ref{t_lines1}, \ref{t_lines3}), but not in $^{13}$CO.
}
\end{table}

\section{Summary}

We report the molecular cloud properties of the N\,55 star forming region in the LMC 
with ALMA observations of $^{12}$CO(1-0) and $^{13}$CO(1-0) emission. This cloud is strongly irradiated by a young star cluster LH\,72 within an expanding SGS, LMC\,4. The results of our analysis 
are summarized as follow:

\begin{enumerate}
  
\item ALMA observations of N\,55 reveal the clumpy nature of CO clouds
  in sub-parsec scales. The cloud properties are analyzed using
  dendrograms which give two sets of clumps, {\it leaves} and {\it
    trunks}. We use both leaves and trunks separately to understand
  their physical properties such as radii, linewidths, luminosities,
  masses, and their relation to star formation. We find that
  molecular cores (dendrogram leaves) that are associated with YSOs
  generally show larger linewidths and masses.
  
\item The $^{12}$CO and $^{13}$CO clump masses are determined by LTE and
virial assumptions. These independent mass estimates show that the LTE
masses of most of the clumps are in very good agreement with the virial masses, and the
LTE versus virial mass plot can be fitted with a power law of
$M_{{\rm VIR}}$=1.75$M_{{\rm LTE}}^{0.68\pm0.12}$, where the virial
ratio is 1.12. The size-linewidth coefficient
($\sigma^2$/R) shows a linear relation with mass surface density as in
many Milky Way and LMC quiescent clouds, where mass surface density is
determined by LTE method. These findings indicate that N\,55 clumps
are in self-gravitational virial equilibrium with negligible external
pressure.

\item The size-linewidth relation is a power law with an index of
  0.5$\pm$0.05, which is consistent with the the size-linewidth
  relation of the Orion cloud, measured at similar spatial scale. This
  result strengthens our argument that N\,55 clouds are
  gravitationally bound and the effects of any external pressure can
  be negligible. We present the size-linewidth relation of $^{12}$CO
  clumps that are identified as dendrogram trunks whose power law
  relation does not show much difference from $^{13}$CO clumps.
\item A power law relation between $^{12}$CO virial mass and
  luminosity is presented, which gives an X$_{{\rm CO}}$ factor of
  6.5$\times$\,10$^{20}$\,cm$^{-2}$\,(K\,km s$^{-1}$)$^{-1}$. This value of X$_{{\rm CO}}$ factor
  is two times the value of Orion cloud, measured for similar spatial scale. 
\item $^{12}$CO and $^{13}$CO core mass functions show a turnover below
  200M$_{\odot}$. This turnover separates
  the steep high mass end from the shallower low-mass part. The low-mass end
  of the $^{12}$CO mass spectrum is fitted with a power law of index
  0.5$\pm$0.1, while for $^{13}$CO it is fitted with a power law index
  0.6$\pm$0.2. The steep
  high-mass end is fitted with a power law index 2.0$\pm$0.3 for $^{13}$CO
  clumps and 2.5$\pm$0.4 for $^{13}$CO. Our studies show that clump mass
  function in N\,55 shows trend similar to Galactic clouds.
\end{enumerate}

\begin{table*}
\centering
\caption{$^{12}$CO(1-0) clump (dendrogram trunks) properties}
\label{t_lines1}
\begin{tabular}{@{}cccccccccccccc}
\hline

&Number ID&   R.A.    &   Decl.    &   Radius (R)&$\delta$R & $\sigma_v$& $\delta\sigma_v$& Velocity     &  \multicolumn{2}{c}{L$_{\rm co}$(K\,km\,s$^{-1}$\,pc$^2$)}&T$_{\rm pk}$ &   M$_{\rm VIR}$&$\delta$M \\
&    &   deg   &   deg    &   pc & pc&km\,s$^{-1}$& km\,s$^{-1}$& km\,s$^{-1}$    & $\delta$L$_{\rm co}$  &$\delta$L$_{\rm co}$ & K &   M$_{\odot}$& M$_{\odot}$\\
\hline
&  1 & 83.1367 & -66.4543 &  3.73 &0.12 & 2.37&0.10 & 288.9  & 2958&2.0 &40.0 & 21824&40 \\        
&  2 & 83.1385 & -66.4603 & 0.80 &0.50 & 1.14&0.50 & 276.3  & 23.8&6.2 &9.1 & 1095&210 \\       
&  3 & 83.0718 & -66.4039 &  1.79 &0.23 &0.86&0.25 & 283.3   & 180.2&3.3 &22.1 & 1389&120 \\       
&  4 & 83.0535 & -66.4027 & 0.94 &0.50 & 1.08&0.45 & 283.7   & 25.3&6.0 &10.2 & 1128& 200\\      
&  5 & 82.9232 & -66.4267 & 0.96 &0.40 &0.602&0.42 & 283.7   & 34.9&6.0 &10.1 & 362&180 \\       
&  6 & 83.0545 & -66.4130 &  4.26&0.10 &  1.58&0.10 & 287.5   & 1082.5&1.4 &25.3 & 11057&45 \\       
&  7 & 82.9230 & -66.4328 & 0.98&0.35 & 0.446&0.30 & 283.7   & 14.2&5.5 &6.9 & 202&90 \\       
&  8 & 83.0979 & -66.4531 &  2.42&0.16 &  1.47&0.20 & 289.4   & 458.1&2.7 &30.2 & 5431&100 \\       
&  9 & 83.0895 & -66.4777 &  1.43&0.35 & 0.97&0.30 & 285.9   & 25.9&4.5 &9.7 & 1416& 120\\       
& 10 & 83.1581 & -66.4704 &  1.07&0.38 &  1.26&0.35 & 286.0   & 33.4&5.5 &10.2 & 1768&135 \\      
& 11 & 83.1484 & -66.4608 & 0.71&0.50 & 0.647&0.50 & 284.5   & 20.1&7.5 &11.4 & 309& 185\\       
& 12 & 83.0427 & -66.4036 &  1.82&0.24 &  1.45&0.23 & 287.4   & 130.6&3.2 &21.6 & 3954& 100\\       
& 13 & 82.9686 & -66.4181 & 0.90&0.50 & 0.46 &0.40 &284.8   & 45.7&5.5 &28.2 & 196&150 \\       
& 14 & 83.1073 & -66.4855 &  1.11&0.43 &  2.08&0.35 & 288.5   & 80.4&8.0 &20.2 & 4983&140 \\       
& 15 & 83.0609 & -66.4384 & 0.81&0.40 &  1.28&0.45 & 286.7   & 21.3&7.0 &10.4 & 1374&170 \\       
& 16 & 83.0404 & -66.4313 &  1.01&0.47 & 0.49&0.30 & 286.2   & 51.7&4.8 &19.5 & 249&95 \\       
& 17 & 83.0547 & -66.4403 & 0.70&0.40 &  1.05&0.52 & 286.5   & 22.4&9.0 &11.4 & 800&200 \\       
& 18 & 82.9795 & -66.4054 &  3.73&0.12 &  1.23&0.12 & 287.9   & 344.7&1.6 &17.1 & 5841&55 \\       
& 19 & 83.0581 & -66.4685 & 0.98&0.30 & 0.61 &0.40 &286.1   & 14.7&5.5 &8.8 & 374&165 \\       
& 20 & 83.1569 & -66.4448 &  1.77&0.30 & 0.79 &0.25 &287.6   & 273.2&3.5 &24.1 & 1137&125 \\       
& 21 & 83.0753 & -66.4467 &  3.16&0.14 &  1.07 &0.10 &288.2   & 258.0&1.4 &13.2 & 3744&40 \\        
& 22 & 83.0266 & -66.4012 &  2.01&0.18 & 0.90 &0.20 &287.3   & 217.3&3.0 &20.9 & 1693&85 \\       
& 23 & 83.0343 & -66.3870 & 0.73&0.40 & 0.66 &0.50 &286.3   & 17.1&7.7 &13.4& 334&190 \\       
& 24 & 83.0904 & -66.4605 &  2.14&0.20 &  1.22&0.20 & 289.1   & 219.2&3.0 &26.7 & 3302&90 \\       
& 25 & 83.1245 & -66.4425 &  1.64&0.25 &  1.99&0.27 & 289.6   & 32.9&4.0 &3.3 & 6775&125 \\       
& 26 & 83.1510 & -66.4679 &  1.74&0.28 &  1.73&0.25 & 288.9   & 23.6&3.4 &2.6 & 5436&115 \\       
& 27 & 83.1618 & -66.4544 &  3.25&0.12 &  1.80&0.13 & 288.5   & 99.6&1.8 &4.3 & 11030&60 \\       
& 28 & 83.1155 & -66.4536 &  1.25&0.38 &  1.26&0.30 & 287.5   & 10.9&5.0 &3.0 & 2068&120 \\       
& 29 & 83.0712 & -66.4402 & 0.97&0.40 & 0.50 &0.40& 286.8      & 17.1&5.5 &15.3 & 259&160 \\       
& 30 & 83.1090 & -66.4331 & 0.93&0.40 & 0.73 &0.45& 287.3    & 16.7&6.5 &5.4 & 521&205 \\       
& 31 & 82.9541 & -66.4189 & 0.95&0.40 & 0.76 &0.35& 286.9    & 17.2&6.5 &6.5 & 576&120 \\       
& 32 & 82.9657 & -66.4131 &  1.38&0.25 & 0.71&0.35 & 287.4   & 71.2&4.0 &16.2 & 722&175 \\       
& 33 & 83.1181 & -66.4829 &  1.60&0.30 &  1.06&0.24 & 288.5   & 122.2&3.7 &18.3 & 1881&100 \\       
& 34 & 83.0252 & -66.4273 &  2.23&0.20 & 0.56&0.18 & 287.2   & 141.3&2.6 &16.4 & 719&75 \\      
& 35 & 83.0171 & -66.4489 &  1.32&0.28 & 0.67&0.28 & 287.8   & 80.5&4.0 &23.1 & 620&110 \\       
& 36 & 82.9924 & -66.4275 & 0.67&0.40 & 0.55&0.30 & 287.2    & 19.6&9.0 &14.9 & 213&65 \\       
& 37 & 82.9958 & -66.4127 & 0.82&0.40 & 0.43&0.30 & 286.8    & 25.1&7.8 &14.6 & 155&80\\       
& 38 & 83.0639 & -66.4578 & 0.95&0.30 & 0.69&0.40 & 288.3    & 56.6&5.5 &18.0 & 473& 160\\       
& 39 & 83.1261 & -66.4460 &  1.16&0.40 &  1.51&0.30 & 288.9    & 9.9&4.8 &3.9 & 2747&110 \\       
& 40 & 83.1401 & -66.4429 &  1.22&0.33 & 0.781&0.30 & 287.4    & 11.3&5.0 &3.3 & 775&115 \\

    \hline
\end{tabular}\\
\end{table*}
%\end{singlespace}
\addtocounter{table}{-1}
%\vspace{-50 in}
\begin{table*}[ht]
\centering
\caption{Continued}
\label{t_lines1}
\begin{tabular}{@{}cccccccccccccc}
\hline

&Number ID&   R.A.    &   Decl.    &   Radius (R)&$\delta$R & $\sigma_v$& $\delta\sigma_v$& Velocity     &  \multicolumn{2}{c}{L$_{\rm co}$(K\,km\,s$^{-1}$\,pc$^2$)}&T$_{\rm pk}$ &   M$_{\rm VIR}$&$\delta$M \\
&    &   deg   &   deg    &   pc &pc &km\,s$^{-1}$&km\,s$^{-1}$ & km\,s$^{-1}$    &$\delta$L$_{\rm co}$   &$\delta$L$_{\rm co}$ & K &   M$_{\odot}$&M$_{\odot}$ \\
\hline
& 41 & 83.1110 & -66.4432 &  1.33&0.30 &  1.12&0.25 & 288.6    & 10.2&4.3 &3.6 & 1724&90 \\       
& 42 & 83.1417 & -66.4321 &  1.16&0.40 & 0.57&0.40 & 288.0    & 48.4&4.6 &19.7 & 392&195 \\       
& 43 & 82.9530 & -66.4161 & 0.94&0.30 & 0.52&0.40 & 287.6    & 32.0&6.7 &17.9 & 265&155 \\       
& 44 & 83.0357 & -66.4396 & 0.94&0.43 & 0.72&0.40 & 288.0    & 10.1&5.0 &5.3 & 505&160 \\       
& 45 & 83.0006 & -66.4211 &  1.85&0.20 &  1.21& 0.23& 290.5    & 174.5&2.8 &19.4 & 2827&100 \\       
& 46 & 83.0744 & -66.4120 &  1.43&0.27 & 0.70&0.29 & 289.3    & 14.3&4.1 &5.8 & 740&125 \\       
& 47 & 83.0312 & -66.4501 &  1.60&0.30 & 0.65& 0.25& 289.3    & 101.5&3.5 &22.8 & 692&105 \\       
& 48 & 83.0659 & -66.4510 & 0.88&0.44 & 0.59&0.40 & 289.6     & 12.9&6.7 &6.5 & 324&150 \\       
& 49 & 83.0189 & -66.3943 &  1.18&0.35 & 0.83&0.30 & 290.2    & 24.4&4.4 &7.2 & 855&110 \\       
& 50 & 83.1110 & -66.4782 &  1.50&0.27 & 0.51&0.30 & 289.9    & 67.2&4.2 &20.6 & 404&140 \\       
& 51 & 83.1126 & -66.4494 & 0.92&0.50 & 0.44&0.35 & 290.0     & 17.7&6.3 &12.5 & 182&117 \\       
& 52 & 83.0756 & -66.4370 &  1.38&0.33 & 0.70 &0.33 & 290.3   & 62.4&3.8 &16.1 & 692&155 \\       
& 53 & 83.1202 & -66.4722 &  1.35&0.28 & 0.45&0.30 & 290.6    & 85.7&4.0 &22.2 & 288&130 \\       
& 54 & 83.1452 & -66.4441 & 0.92&0.48 & 0.63&0.40 & 290.8    & 29.8&7.2 &13.5 & 385&153 \\       
& 55 & 83.1504 & -66.4502 &  1.07&0.33 & 0.81&0.40 & 291.1    & 26.1&5.2 &4.5 & 729&177 \\       
& 56 & 83.0355 & -66.4186 & 0.91&0.50 & 0.91&0.42 & 293.2    & 14.1&6.0 &4.5 & 781&167 \\       
& 57 & 83.0286 & -66.4206 & 0.88&0.50 & 0.54&0.46 & 293.7    & 31.0&6.8 &21.3 & 268&195 \\ 

    \hline
\end{tabular}\\
\end{table*}
%\newpage
%\addtocounter{table}
\begin{table*}
  \addtolength{\tabcolsep}{-3pt}
\centering
\caption{$^{13}$CO(1-0) clump (dendrogram trunks) properties}
\label{t_lines2}
\begin{tabular}{@{}cccccccccccccccccc}
\hline
%     & RA      &     Dec     &     Radius  &     Flux     &     Velocity dispersion & Luminosity& Virial mass \\
%              &             &             &              &          $\sigma$         & L$_{co}$     & M$_\odot$    \\
%\hline
& Number ID&  R.A.  & Decl. & Radius (R) &$\delta$R &$\sigma_v$ &$\delta\sigma_v$ & Velocity & T$_{\rm pk}$ &\multicolumn{2}{c}{L$_{\rm co}$}(K\,km\,s$^{-1}$\,pc$^2$) &N$_{\rm H_2}$  & M$_{\rm LTE}$ & $\delta$M$_{\rm LTE}$ & M$_{\rm VIR}$&$\delta$M  \\
&       &  deg   & dec& pc     &pc      & km\,s$^{-1}$ &   km\,s$^{-1}$         &   km\,s$^{-1}$&K & $\delta$L$_{\rm co}$   & $\delta$L$_{\rm co}$   &10$^{21}$cm$^{-2}$    & M$_{\odot}$&M$_{\odot}$  & M$_{\odot}$& M$_{\odot}$ \\
\hline
&  1 & 83.1354 & -66.4586 & 0.55 &0.30 &0.58&0.40 & 282.0 &1.99&1.7 &0.5 &4.0 & 171 &50 &191& 90\\
&  2 & 83.0713 & -66.4044 &  1.23 &0.35 &0.65&0.40 & 283.1&1.38& 9.5&0.2 &5.0 & 967 &23 &532&200 \\
&  3 & 83.1374 & -66.4563 & 0.65 &0.30 &0.48&0.30 & 282.8 &1.15& 1.3&0.4 &2.9 & 136 &45 &156&60 \\
&  4 & 82.9227 & -66.4325 & 0.80 &0.40 &0.29&0.20 & 283.5 &1.96& 1.1& 0.3&1.0 & 66 &20 &72& 35\\
&  5 & 83.0486 & -66.4152 & 0.63 &0.30 &0.53&0.30 & 284.8 &2.16&2.4 &0.5 &3.8 & 245 &45 &183&60 \\
&  6 & 83.0623 & -66.4126 &  2.33 &0.20 & 1.16&0.45 & 287.2 &3.77&40 &0.1 &9.9 & 5097 &15 &3237& 490\\
&  7 & 83.1374 & -66.4539 &  2.98 &0.15 & 1.42&0.42 & 288.9 &5.39&154 &0.1 &27 & 28278 &20 &6222& 550\\
&  8 & 82.9683 & -66.4180 & 0.73 &0.40 &0.36&0.28 & 284.9 &2.74&2.3 &0.3 &4.4 & 305 &45 &98& 60\\
&  9 & 82.9886 & -66.4064 & 0.55 &0.30 &0.39&0.25 & 286.4 &1.94&1.2 &0.5 &3.3 & 123 &55 &85& 35\\
& 10 & 83.0404 & -66.4313 & 0.88 &0.40 &0.39& 0.25& 286.2 &3.03&4.1 &0.3 &4.1 & 443 &34 &142& 60\\
& 11 & 83.1567 & -66.4451 &  1.55 &0.30 &0.65& 0.38& 287.6&4.20&19.3 &0.2 &8.5 & 2357 &20 &680& 235\\
& 12 & 82.9783 & -66.4053 &  1.04 &0.40 &0.55&0.40 & 287.1&1.72&1.6 & 0.3&1.7 & 149 &25 &322& 172\\
& 13 & 82.9644 & -66.4129 & 0.75 &0.40 &0.88&0.45 & 287.3 &0.94&1.4 &0.4 &2.0 & 124 &35 &602& 160\\
& 14 & 83.0264 & -66.4017 &  1.44 &0.30 &0.53& 0.40& 287.2&1.91&9.7 &0.2 &4.8 & 969 &25 &414&240 \\
& 15 & 82.9956 & -66.4127 & 0.69 &0.35 &0.31&0.25 & 286.7 &1.54&1.1 &0.5 &1.8 & 97 &44 &71& 45\\
& 16 & 83.1187 & -66.4828 &  1.20 &0.35 &0.76&0.40 & 288.2&1.52&4.0 &0.2 &2.9 & 374 &22 &722&200 \\
& 17 & 83.0953 & -66.4526 &  1.36 &0.40 &0.91&0.40 & 288.8&6.58&23.0 &0.2 &14 & 3662 &30 &1162&225 \\
& 18 & 83.0642 & -66.4441 & 0.81 &0.40 &0.52&0.40 & 287.3 &3.11&2.6 &0.4 & 3.8 & 319 &47 &225& 135\\
& 19 & 82.9715 & -66.4041 & 0.61 &0.30 &0.57&0.38 & 287.3 &1.10&1.0 &0.4 & 1.9 & 85 &33 &205& 92\\
& 20 & 83.0251 & -66.4268 &  1.65 &0.28 &0.47&0.35 & 287.3&1.29&4.5 &0.2 & 2.3 & 422 &18 &377& 210\\
& 21 & 83.0435 & -66.4136 &  1.25 &0.30 &0.87& 0.42& 288.6&3.80&19.0 &0.2 & 13 & 2835 &30 &995& 230\\
& 22 & 83.1243 & -66.4558 & 0.54 &0.30 &0.55&0.40 & 287.6 &2.47&2.2 &0.7 &  5.6 & 251 &78 &173& 90\\
& 23 & 83.0425 & -66.4038 & 0.81 &0.40 &0.83& 0.40& 288.3 &1.13&1.7 &0.5 &2.2 & 173 &50 &586&135 \\
& 24 & 83.0930 & -66.4601 &  1.08&0.40 & 0.86& 0.50& 289.2&2.73&7.0 &0.3 &6.7 & 936 &40 &836&280 \\
& 25 & 83.0640 & -66.4576 & 0.75&0.40 & 0.47& 0.40& 288.4 &2.25&2.4 &0.4 &3.2 & 238 &40 &173&125 \\
& 26 & 82.9536 & -66.4163 & 0.67&0.32 & 0.43&0.30 & 287.8 &1.08&0.9 &0.4 &1.5 & 81 &35 &128&65 \\
& 27 & 83.1077 & -66.4858 & 0.67&0.32 & 0.57&0.40 & 289.4 &2.32&2.4 &0.4 &4.1 & 278 &45 &225&112 \\
& 28 & 83.0165 & -66.4488 & 0.85&0.40 & 0.37&0.25 & 288.2 &1.80&2.0 &0.3 &2.4 & 214 &30 &125& 55\\
& 29 & 83.0813 & -66.4489 &  1.00&0.52 & 0.48&0.30 & 288.9 &2.77&5.7 &0.3 &5.7 & 738 &35 &246&95 \\
& 30 & 83.1419 & -66.4321 & 0.80&0.40 & 0.38&0.25 & 288.2 &2.49&2.4 &0.4 &3.1 & 246 &35 &120& 52\\
& 31 & 83.0466 & -66.4090 & 0.53&0.30 & 0.49&0.28 & 288.3 &1.28&0.9 &0.6 &2.1 & 76 &45 &131& 45\\
& 32 & 83.0714 & -66.4469 & 0.99&0.40 & 0.55&0.30 & 288.8 &1.57&2.5 &0.3 &2.2 & 212 &22 &307& 95\\
& 33 & 83.0312 & -66.4501 &  1.41&0.32 & 0.47& 0.28& 289.4&2.35&3.7 &0.2 &3.1 & 441 &25 &319&115 \\

\hline
\end{tabular}
\end{table*}

\addtocounter{table}{-1}

\begin{table}
\addtolength{\tabcolsep}{-3pt}
\centering
\caption{Continued}
\label{t_lines2}
\begin{tabular}{@{}ccccccccccccccccc}
\hline
%     & RA      &     Dec     &     Radius  &     Flux     &     Velocity dispersion & Luminosity& Virial mass \\
%              &             &             &              &          $\sigma$         & L$_{co}$     & M$_\odot$    \\
%\hline
& Number ID&  R.A.  & Decl. & Radius (R) &$\delta$R &$\sigma_v$ &$\delta\sigma_v$ & Velocity & T$_{\rm pk}$ &\multicolumn{2}{c}{L$_{\rm co}$}(K\,km\,s$^{-1}$\,pc$^2$) &N$_{\rm H_2}$  & M$_{\rm LTE}$ & $\delta$M &M$_{\rm VIR}$&$\delta$M  \\
&       &  deg   & dec& pc     &pc      & km\,s$^{-1}$ &   km\,s$^{-1}$         &   km\,s$^{-1}$&K &    & $\delta$L$_{\rm co}$   &10$^{21}$cm$^{-2}$    & M$_{\odot}$  &M$_{\odot}$ & M$_{\odot}$& M$_{\odot}$ \\
\hline
& 34 & 82.9848 & -66.4055 & 0.97&0.45 & 0.38&0.25 & 289.2 &2.86&3.5 &0.3 &3.2 & 359 &28 &144& 65\\
& 35 & 82.9704 & -66.4098 & 0.72&0.35 & 0.37&0.25 & 289.3 &1.60&1.0 &0.3 &1.6 & 81 &27 &102& 50\\
& 36 & 83.1375 & -66.4616 & 0.97&0.45 & 0.80&0.42 & 290.3 &3.79&6.8 &0.4 &7.6 & 1083 &62 &642& 180\\
& 37 & 83.0014 & -66.4205 &  1.11&0.43 &  1.01&0.40 & 290.5 &2.59&5.5 &0.2 &4.0 & 544 &24 &1188&190 \\
& 38 & 83.0770 & -66.4365 & 0.68&0.37 & 0.61&0.30 & 290.2 &1.40&1.3 &0.3 &2.2 & 116 &30 &264&65 \\
& 39 & 82.9953 & -66.4220 & 0.64&0.30 & 0.36&0.25 & 290.2 &1.82&1.0 &0.5 &1.9 & 85 &42 &86& 40\\
& 40 & 83.1098 & -66.4774 & 0.73&0.35 & 0.25&0.20 & 290.2 &2.10&1.5 &0.4 &2.5 & 161 &41 &46& 30\\
& 41 & 83.1210 & -66.4716 & 0.82&0.40 & 0.36&0.28 & 290.5 &2.45&2.5 &0.4 &3.6 & 310 &46 &108& 70\\
& 42 & 83.1453 & -66.4443 & 0.69&0.30 & 0.34&0.25 & 290.9 &1.80&1.1 &0.4 &1.8 & 87 &30 &82& 45\\
& 43 & 83.1079 & -66.4542 & 0.99&0.45 & 0.60&0.30 & 291.4 &0.93&1.2 &0.3 &1.3 & 95 &23 &365&95\\

\hline
\end{tabular}
\end{table}
\newpage

%{\bf Online only} \\

\begin{table*}
\centering
\caption{$^{13}$CO(1-0) core (dendrogram leaves) properties}
\label{t_lines4}
\begin{tabular}{@{}ccccccccccccccccc}
\hline
%     & RA      &     Dec     &     Radius  &     Flux     &     Velocity dispersion & Luminosity& Virial mass \\
%              &             &             &              &          $\sigma$         & L$_{co}$     & M$_\odot$    \\
%\hline
& Number ID&  R.A.  & Decl. & Radius &$\sigma_v$ & Velocity & T$_{\rm pk}$ & L$_{\rm co}$  & M$_{\rm LTE}$ &  M$_{\rm VIR}$\\

&       &  deg   & dec& pc       & km\,s$^{-1}$          &   km\,s$^{-1}$&K &  K\,km\,s$^{-1}$\,pc$^2$      & M$_{\odot}$  & M$_{\odot}$ \\
\hline

&  1 & 83.1354 & -66.4586 & 0.55 & 0.58 & 282.0 & 2.0 & 1.7 &  160 & 191 \\
&  2 & 83.1374 & -66.4563 & 0.65 & 0.48 & 282.8 & 1.2 & 1.3 &  135 & 156 \\
&  3 & 83.0712 & -66.4042 & 0.76 & 0.43 & 283.2 & 1.6 & 3.1 &  312 & 145 \\
%&  4 & 83.0697 & -66.4027 & 0.29 & 0.09 & 282.3 & 1.5 & 0.2 &   14 &   3 \\
&  4 & 82.9227 & -66.4325 & 0.80 & 0.29 & 283.6 & 1.4 & 1.1 &   62 &  71 \\
&  5 & 83.0486 & -66.4152 & 0.63 & 0.53 & 284.8 & 2.2 & 2.4 &  245 & 183 \\
&  6 & 82.9683 &  -66.418 & 0.73 & 0.36 & 284.9 & 2.7 & 2.3 &  305 &  98 \\
&  7 & 83.1357 & -66.4546 & 0.59 & 0.14 & 284.9 & 4.4 & 0.2 &   41 &  13 \\
&  8 & 83.0605 & -66.4122 & 1.14 & 0.57 & 286.2 & 3.1 & 7.7 &  973 & 386 \\
& 9 & 82.9886 & -66.4064 & 0.55 & 0.39 & 286.4 & 1.9 & 1.2 &  123 &  85 \\
& 10 & 83.1449 &  -66.455 & 0.66 & 0.28 & 285.8 & 1.9 & 1.2 &  112 &  54 \\
%& 11 & 83.1293 & -66.4519 & 0.40 & 0.17 & 286.0 & 1.5 & 0.2 &   23 &  12 \\
& 11 & 83.0411 & -66.4317 & 0.46 & 0.23 & 286.0 & 3.0 & 1.3 &  137 &  26 \\
& 12 & 82.9644 & -66.4129 & 0.75 & 0.88 & 287.3 & 0.9 & 1.4 &  125 & 602 \\
& 13 & 83.0594 & -66.4094 & 0.81 & 0.43 & 286.5 & 1.8 & 2.5 &  216 & 153 \\
& 14 & 82.9956 & -66.4127 & 0.69 & 0.31 & 286.8 & 1.5 & 1.1 &   97 &  71 \\
& 15 & 83.1187 & -66.4828 & 1.20 & 0.76 & 288.2 & 1.5 & 4.0 &  374 & 722 \\
& 16 & 83.0642 & -66.4441 & 0.81 & 0.52 & 287.3 & 2.8 & 2.6 &  317 & 225 \\
& 17 & 83.0226 & -66.4281 & 1.01 & 0.36 & 287.2 & 1.0 & 1.6 &  128 & 136 \\
& 18 & 82.9715 & -66.4041 & 0.61 & 0.57 & 287.3 & 1.1 & 1.0 &   85 & 205 \\
%& 21 & 83.0391 & -66.4304 & 0.33 & 0.17 & 286.4 & 2.2 & 0.4 &   42 &  10 \\
& 19 & 83.1243 & -66.4558 & 0.54 & 0.55 & 287.6 & 2.5 & 2.2 &  251 & 173 \\
& 20 &  83.027 & -66.4258 & 0.88 & 0.41 & 287.1 & 1.8 & 2.0 &  189 & 153 \\
& 21 & 82.9772 & -66.4049 & 0.62 & 0.29 & 287.2 & 1.1 & 0.7 &   60 &  52 \\
& 22 & 83.0278 & -66.4023 & 0.60 & 0.34 & 287.1 & 2.1 & 1.2 &  123 &  74 \\
& 23 & 83.0239 & -66.4001 & 0.56 & 0.47 & 287.3 & 3.5 & 2.8 &  328 & 129 \\
& 24 & 83.1566 & -66.4446 & 0.70 & 0.38 & 287.4 & 3.6 & 4.3 &  513 & 107 \\
& 25 & 83.0435 & -66.4139 & 0.72 & 0.58 & 288.2 & 3.6 & 8.1 & 1205 & 250 \\
& 26 & 82.9802 & -66.4059 & 0.66 & 0.22 & 286.9 & 1.5 & 0.5 &   44 &  34 \\
& 27 & 83.0303 & -66.4044 & 0.52 & 0.42 & 287.2 & 3.6 & 2.1 &  228 &  97 \\
& 28 & 83.0425 & -66.4038 & 0.81 & 0.83 & 288.3 & 1.1 & 1.7 &  173 & 586 \\
& 29 &  83.064 & -66.4576 & 0.75 & 0.47 & 288.4 & 2.0 & 2.4 &  236 & 173 \\
& 30 & 83.0949 & -66.4523 & 0.54 & 0.67 & 288.5 & 6.6 & 9.2 & 1482 & 253 \\
& 31 & 82.9536 & -66.4163 & 0.67 & 0.43 & 287.7 & 1.2 & 0.9 &   81 & 128 \\
%& 35 & 83.1461 & -66.4578 & 0.29 & 0.21 & 287.7 & 2.0 & 0.3 &   31 &  13 \\
%& 33 & 83.1361 & -66.4546 & 0.40 & 0.36 & 287.7 & 4.4 & 2.5 &  391 &  55 \\
& 32 & 83.1077 & -66.4858 & 0.67 & 0.57 & 289.4 & 1.7 & 2.4 &  274 & 225 \\
%& 35 & 83.1372 & -66.4524 & 0.37 & 0.53 & 288.3 & 5.2 & 2.1 &  360 & 106 \\
& 33 & 83.0165 & -66.4488 & 0.86 & 0.37 & 288.2 & 1.4 & 2.0 &  211 & 125 \\

\hline
\end{tabular}\\
\end{table*}

\addtocounter{table}{-1}

\begin{table*}
%\addtocounter{table}
\centering
\caption{$^{13}$CO(1-0) core (dendrogram leaves) properties}
\label{t_lines4}
\begin{tabular}{@{}ccccccccccccccccc}
\hline
%     & RA      &     Dec     &     Radius  &     Flux     &     Velocity dispersion & Luminosity& Virial mass \\
%              &             &             &              &          $\sigma$         & L$_{co}$     & M$_\odot$    \\
%\hline
& Number ID&  R.A.  & Decl. & Radius &$\sigma_v$ & Velocity & T$_{\rm pk}$ & L$_{\rm co}$  & M$_{\rm LTE}$ &  M$_{\rm VIR}$\\

&       &  deg   & dec& pc       & km\,s$^{-1}$          &   km\,s$^{-1}$&K &  K\,km\,s$^{-1}$\,pc$^2$      & M$_{\odot}$  & M$_{\odot}$ \\
\hline
& 34 & 83.0811 & -66.4488 & 0.91 & 0.48 & 288.9 & 2.8 & 5.3 &  679 & 220 \\
& 35 & 83.1419 & -66.4321 & 0.80 & 0.38 & 288.2 & 2.5 & 2.4 &  246 & 120 \\
& 36 & 83.0466 &  -66.409 & 0.53 & 0.49 & 288.3 & 1.2 & 1.0 &   76 & 131 \\
& 37 & 83.1005 & -66.4526 & 0.49 & 0.41 & 288.7 & 2.4 & 1.6 &  170 &  84 \\
& 38 & 83.1561 & -66.4467 & 0.42 & 0.24 & 288.0 & 3.5 & 0.8 &   82 &  25 \\
& 39 & 83.0623 & -66.4131 & 0.63 & 0.36 & 288.3 & 2.9 & 2.3 &  274 &  86 \\
& 40 & 83.1344 & -66.4591 & 0.76 & 0.44 & 288.8 & 3.3 & 3.3 &  423 & 155 \\
& 41 & 83.1431 & -66.4563 & 0.45 & 0.29 & 288.3 & 3.8 & 1.6 &  178 &  39 \\
& 42 & 83.1275 &  -66.454 & 0.88 & 0.37 & 288.6 & 2.8 & 3.9 &  405 & 124 \\
& 43 & 83.1311 & -66.4526 & 0.45 & 0.17 & 288.0 & 2.6 & 0.6 &   69 &  14 \\
& 44 & 83.0729 & -66.4471 & 0.70 & 0.43 & 288.5 & 1.6 & 1.2 &  104 & 137 \\
& 45 & 83.0673 & -66.4149 & 0.54 & 0.28 & 288.4 & 5.0 & 3.5 &  500 &  44 \\
%& 52 & 83.0947 & -66.4617 & 0.34 & 0.29 & 288.4 & 1.5 & 0.3 &   31 &  30 \\
& 46 & 83.0936 & -66.4602 & 0.60 & 0.78 & 289.3 & 2.9 & 3.8 &  516 & 384 \\
& 47 & 82.9848 & -66.4055 & 0.97 & 0.38 & 289.2 & 2.9 & 3.5 &  359 & 144 \\
& 48 & 83.1404 &  -66.455 & 0.64 & 0.42 & 289.0 & 3.9 & 5.6 &  742 & 117 \\
& 49 & 82.9704 & -66.4098 & 0.72 & 0.37 & 289.3 & 1.6 & 1.0 &   81 & 102  \\
& 50 & 83.0312 & -66.4493 & 0.78 & 0.42 & 289.3 & 2.4 & 2.1 &  253 & 145 \\
& 51 & 83.1406 & -66.4617 & 0.53 & 0.22 & 289.1 & 1.4 & 0.5 &   43 &  27 \\
%& 59 & 83.0899 & -66.4589 & 0.39 & 0.30 & 289.3 & 1.3 & 0.4 &   33 &  37 \\
& 52 & 83.1446 & -66.4476 & 0.51 & 0.36 & 289.3 & 1.6 & 0.9 &   77 &  70 \\
%& 61 & 83.1391 & -66.4514 & 0.34 & 0.27 & 289.3 & 3.6 & 0.9 &  142 &  26 \\
%& 62 & 83.0868 & -66.4506 & 0.37 & 0.15 & 289.1 & 0.9 & 0.2 &   14 &   8 \\
& 53 &  83.031 & -66.4516 & 0.60 & 0.20 & 289.3 & 2.1 & 0.5 &   46 &  24 \\
& 54 & 83.0686 & -66.4466 & 0.50 & 0.27 & 289.4 & 1.2 & 0.5 &   33 &  37 \\
& 55 &  83.077 & -66.4365 & 0.68 & 0.61 & 290.2 & 1.2 & 1.3 &  115 & 264 \\
%& 66 & 83.0045 & -66.4207 & 0.38 & 0.34 & 289.5 & 2.0 & 0.6 &   59 &  45 \\
%& 67 & 83.0448 & -66.4124 & 0.25 & 0.11 & 289.3 & 3.2 & 0.2 &   26 &   3 \\
& 56 & 83.1372 & -66.4616 & 0.85 & 0.69 & 290.4 & 3.8 & 5.5 &  876 & 424 \\
& 57 & 82.9953 &  -66.422 & 0.64 & 0.36 & 290.2 & 1.8 & 1.0 &   85 &  86 \\
& 58 & 83.0946 & -66.4535 & 0.47 & 0.14 & 289.8 & 2.4 & 0.4 &   50 &  10 \\
%& 71 & 83.0323 & -66.4531 & 0.37 & 0.23 & 289.9 & 1.1 & 0.3 &   25 &  21 \\
& 59 & 82.9999 & -66.4202 & 0.48 & 0.52 & 290.8 & 2.6 & 1.9 &  193 & 135 \\
& 60 & 83.1098 & -66.4774 & 0.73 & 0.25 & 290.2 & 2.1 & 1.5 &  161 &  46 \\
& 61 &  83.121 & -66.4716 & 0.82 & 0.36 & 290.5 & 2.3 & 2.5 &  313 & 108 \\
%& 66 & 83.1357 & -66.4529 & 0.40 & 0.27 & 289.9 & 5.2 & 1.9 &  333 &  30 \\
& 62 & 83.1382 &  -66.457 & 0.47 & 0.32 & 290.4 & 3.3 & 1.7 &  231 &  51 \\
%& 68 & 83.1352 & -66.4547 & 0.37 & 0.39 & 290.4 & 3.9 & 1.7 &  268 &  59 \\
%& 78 & 83.0456 & -66.4121 & 0.29 & 0.10 & 289.9 & 3.2 & 0.3 &   42 &   3 \\
& 63 & 83.1453 & -66.4443 & 0.69 & 0.34 & 290.9 & 1.4 & 1.0 &   86 &  82 \\
& 64 & 83.1079 & -66.4542 & 0.99 & 0.60 & 291.4 & 0.7 & 1.2 &   81 & 365 \\
%& 81 & 82.9984 & -66.4206 & 0.33 & 0.19 & 292.9 & 1.2 & 0.2 &   16 &  12 \\

\hline
\end{tabular}\\
\end{table*}
\begin{table*}[h]
\addtolength{\tabcolsep}{2pt}
%\addtocounter{table}
\centering
\caption{$^{12}$CO(1-0) core (dendrogram leaves) properties}
\label{t_lines3}
\begin{tabular}{@{}ccccccccccccccccc}
\hline
%     & RA      &     Dec     &     Radius  &     Flux     &     Velocity dispersion & Luminosity& Virial mass \\
%              &             &             &              &          $\sigma$         & L$_{co}$     & M$_\odot$    \\
%\hline
& Number ID&  R.A.  & Decl. & Radius &$\sigma_v$ & Velocity & L$_{\rm co}$ & M$_{\rm VIR}$\\

&       &  deg   & dec& pc       & km\,s$^{-1}$          &   km\,s$^{-1}$&  K\,km\,s$^{-1}$\,pc$^2$      &  M$_{\odot}$ \\
\hline
%&   1 & 83.0412 & -66.4040 & 0.40 & 0.31 & 289.3 &   4.9 &   39 \\
%&   2 & 83.0491 & -66.4148 & 0.40 & 0.25 & 286.9 &   2.2 &   25 \\
%&   3 & 83.1190 & -66.4744 & 0.40 & 0.14 & 290.8 &   2.0 &    8 \\
%&   4 & 83.0752 & -66.4413 & 0.40 & 0.17 & 286.5 &   1.0 &   12 \\
%&   5 & 82.9863 & -66.4016 & 0.40 & 0.23 & 289.3 &   1.6 &   23 \\
%&   6 & 83.1402 & -66.4326 & 0.41 & 0.11 & 286.4 &   0.5 &    5 \\
%&   7 & 83.0885 & -66.4602 & 0.41 & 0.32 & 291.4 &   3.9 &   42 \\
%&   8 & 83.0650 & -66.4453 & 0.41 & 0.29 & 289.4 &   1.1 &   35 \\
%&   9 & 83.0949 & -66.4617 & 0.41 & 0.33 & 288.1 &  10.8 &   47 \\
%&  10 & 83.0834 & -66.4509 & 0.41 & 0.20 & 289.3 &   2.6 &   18 \\
%&  11 & 83.0893 & -66.4785 & 0.41 & 0.16 & 284.6 &   0.9 &   11 \\
%&  12 & 82.9721 & -66.4041 & 0.42 & 0.16 & 287.2 &   4.8 &   12 \\
%&  13 & 83.0217 & -66.4508 & 0.42 & 0.19 & 287.3 &   4.1 &   16 \\
%&  14 & 83.1118 & -66.4826 & 0.42 & 0.22 & 290.3 &   3.5 &   22 \\
&  1 & 83.0709 & -66.4376 & 0.42 & 0.32 & 290.9 &   4.5 &   43 \\
&  2 & 83.0410 & -66.4317 & 0.42 & 0.20 & 286.0 &   8.3 &   18 \\
&  3 & 82.9905 & -66.4059 & 0.42 & 0.28 & 288.3 &   3.1 &   35 \\
&  4 & 83.0732 & -66.4465 & 0.42 & 0.20 & 289.1 &   5.6 &   18 \\
&  5 & 83.0846 & -66.4541 & 0.43 & 0.31 & 291.3 &   4.0 &   43 \\
&  6 & 82.9798 & -66.4027 & 0.43 & 0.22 & 286.9 &   5.7 &   22 \\
&  7 & 83.0610 & -66.4387 & 0.43 & 0.42 & 287.3 &   6.5 &   80 \\
&  8 & 82.9762 & -66.4045 & 0.44 & 0.16 & 288.4 &   2.5 &   11 \\
&  9 & 83.0378 & -66.4024 & 0.44 & 0.49 & 285.1 &   7.6 &  109 \\
&  10 & 83.1184 & -66.4756 & 0.44 & 0.34 & 290.0 &   2.6 &   52 \\
&  11 & 83.1245 & -66.4494 & 0.44 & 0.46 & 289.0 &   3.2 &   95 \\
&  12 & 83.0782 & -66.4629 & 0.45 & 0.42 & 286.4 &   0.9 &   83 \\
&  13 & 83.1252 & -66.4477 & 0.45 & 0.38 & 287.7 &   2.1 &   69 \\
&  14 & 83.1365 & -66.4612 & 0.45 & 0.66 & 279.1 &   2.1 &  202 \\
&  15 & 83.1297 & -66.4519 & 0.45 & 0.36 & 287.6 &  10.5 &   60 \\
&  16 & 83.0772 & -66.4122 & 0.46 & 0.41 & 289.1 &   3.6 &   81 \\
&  17 & 83.0214 & -66.3953 & 0.46 & 0.26 & 290.4 &   2.2 &   34 \\
&  18 & 82.9904 & -66.4070 & 0.47 & 0.11 & 288.9 &   1.8 &    6 \\
&  19 & 83.0262 & -66.3988 & 0.47 & 0.20 & 288.4 &   3.5 &   20 \\
&  20 & 83.0750 & -66.4460 & 0.47 & 0.28 & 287.4 &   3.5 &   38 \\
&  21 & 83.1300 & -66.4569 & 0.47 & 0.23 & 290.3 &   3.3 &   26 \\
&  22 & 83.0273 & -66.4259 & 0.48 & 0.25 & 287.1 &   8.7 &   30 \\
&  23 & 83.0467 & -66.4091 & 0.48 & 0.36 & 288.5 &   9.1 &   65 \\
&  24 & 82.9768 & -66.4050 & 0.48 & 0.16 & 287.2 &   4.5 &   13 \\
&  25 & 83.1566 & -66.4442 & 0.48 & 0.15 & 287.3 &   7.7 &   12 \\
&  26 & 82.9703 & -66.4101 & 0.48 & 0.20 & 289.1 &   6.8 &   20 \\
&  27 & 82.9705 & -66.4077 & 0.49 & 0.13 & 289.6 &   1.9 &    8 \\
&  28 & 83.0187 & -66.3941 & 0.51 & 0.53 & 290.1 &   6.5 &  146 \\
&  29 & 83.1476 & -66.4541 & 0.51 & 0.31 & 293.3 &   3.0 &   50 \\
&  30 & 83.0769 & -66.4101 & 0.51 & 0.26 & 288.2 &   1.2 &   36 \\

\hline
\end{tabular}\\
\end{table*}
%\FloatBarrier
%\section{Comparison with star formation}
\addtocounter{table}{-1}

\begin{table*}
 \addtolength{\tabcolsep}{2pt}
%\addtocounter{table}
\centering
\caption{Continued}
\label{t_lines3}
\begin{tabular}{@{}ccccccccccccccccc}
\hline
%     & RA      &     Dec     &     Radius  &     Flux     &     Velocity dispersion & Luminosity& Virial mass \\
%              &             &             &              &          $\sigma$         & L$_{co}$     & M$_\odot$    \\
%\hline
& Number ID&  R.A.  & Decl. & Radius &$\sigma_v$ & Velocity  L$_{\rm co}$ & M$_{\rm VIR}$\\

&       &  deg   & dec& pc       & km\,s$^{-1}$          &   km\,s$^{-1}$&  K\,km\,s$^{-1}$\,pc$^2$      &  M$_{\odot}$ \\
\hline
&  31 & 83.1064 & -66.4539 & 0.51 & 0.22 & 290.8 &   3.8 &   26 \\
&  32 & 83.1480 & -66.4495 & 0.51 & 0.56 & 290.8 &  12.3 &  167 \\
&  33 & 83.0706 & -66.4403 & 0.51 & 0.46 & 286.8 &  12.1 &  115 \\
&  34 & 83.1296 & -66.4591 & 0.52 & 0.71 & 277.7 &   8.1 &  275 \\
&  35 & 83.1252 & -66.4533 & 0.52 & 0.27 & 288.9 &  16.0 &   40 \\
&  36 & 82.9949 & -66.4221 & 0.53 & 0.32 & 290.4 &  11.3 &   57 \\
&  37 & 83.1717 & -66.4567 & 0.53 & 0.59 & 289.8 &   1.0 &  194 \\
&  38 & 83.1133 & -66.4832 & 0.54 & 0.29 & 287.2 &   6.2 &   47 \\
&  39 & 83.1260 & -66.4588 & 0.54 & 0.54 & 275.4 &   3.4 &  168 \\
&  40 & 83.1355 & -66.4584 & 0.54 & 0.63 & 281.8 &  19.3 &  226 \\
&  41 & 83.1244 & -66.4557 & 0.55 & 0.63 & 287.8 &  31.8 &  222 \\
&  42 & 83.0906 & -66.4786 & 0.55 & 0.49 & 286.6 &  11.3 &  138 \\
&  43 & 83.0760 & -66.4007 & 0.55 & 0.22 & 284.5 &   2.7 &   28 \\
&  44 & 83.1346 & -66.4538 & 0.55 & 0.47 & 289.1 &  26.4 &  124 \\
&  45 & 83.0307 & -66.4515 & 0.56 & 0.25 & 289.4 &   9.8 &   35 \\
&  46 & 83.0613 & -66.4128 & 0.56 & 0.31 & 287.9 &  15.1 &   58 \\
&  47 & 83.1541 & -66.4515 & 0.56 & 0.52 & 292.3 &   3.2 &  158 \\
&  48 & 83.0531 & -66.4027 & 0.56 & 0.60 & 283.4 &   9.9 &  210 \\
&  49 & 83.0214 & -66.4317 & 0.57 & 0.23 & 287.0 &   6.8 &   33 \\
&  50 & 83.1523 & -66.4506 & 0.58 & 0.40 & 290.9 &   3.8 &   95 \\
&  51 & 83.1399 & -66.4550 & 0.58 & 0.31 & 289.4 &  23.7 &   56 \\
&  52 & 83.1451 & -66.4551 & 0.58 & 0.74 & 285.4 &  27.6 &  327 \\
&  53 & 83.1124 & -66.4799 & 0.58 & 0.23 & 289.5 &   9.6 &   33 \\
&  54 & 83.0822 & -66.4627 & 0.58 & 0.44 & 289.3 &  14.4 &  115 \\
&  55 & 83.0447 & -66.4014 & 0.59 & 0.59 & 286.0 &  13.5 &  212 \\
&  56 & 83.0487 & -66.4152 & 0.60 & 0.57 & 285.0 &  28.7 &  198 \\
&  57 & 83.0554 & -66.4025 & 0.61 & 0.39 & 286.1 &   1.9 &   97 \\
&  58 & 82.9965 & -66.4089 & 0.61 & 0.28 & 288.3 &   6.4 &   49 \\
&  59 & 83.0715 & -66.4125 & 0.61 & 0.43 & 289.7 &   4.8 &  116 \\
&  60 & 83.0279 & -66.3982 & 0.62 & 0.42 & 285.8 &   9.1 &  115 \\
&  61 & 83.1100 & -66.4773 & 0.62 & 0.27 & 290.2 &  20.5 &   46 \\
&  62 & 82.9720 & -66.4143 & 0.62 & 0.32 & 287.7 &   6.2 &   66 \\
&  63 & 83.1611 & -66.4535 & 0.63 & 0.32 & 288.7 &   2.3 &   65 \\
&  64 & 83.1113 & -66.4556 & 0.63 & 0.63 & 290.6 &  19.3 &  264 \\
&  65 & 83.0429 & -66.4041 & 0.63 & 0.42 & 287.6 &  25.8 &  119 \\

\hline
\end{tabular}\\
\end{table*}

\addtocounter{table}{-1}

\begin{table*}
 \addtolength{\tabcolsep}{2pt}
%\addtocounter{table}
\centering
\caption{Continued}
\label{t_lines3}
\begin{tabular}{@{}ccccccccccccccccc}
\hline
%     & RA      &     Dec     &     Radius  &     Flux     &     Velocity dispersion & Luminosity& Virial mass \\
%              &             &             &              &          $\sigma$         & L$_{co}$     & M$_\odot$    \\
%\hline
& Number ID&  R.A.  & Decl. & Radius &$\sigma_v$ & Velocity  L$_{\rm co}$ & M$_{\rm VIR}$\\

&       &  deg   & dec& pc       & km\,s$^{-1}$          &   km\,s$^{-1}$&  K\,km\,s$^{-1}$\,pc$^2$      &  M$_{\odot}$ \\
\hline
&  66 & 83.0573 & -66.4128 & 0.64 & 0.24 & 286.6 &  20.9 &   38 \\
&  67 & 82.9879 & -66.4064 & 0.64 & 0.32 & 286.4 &  21.4 &   70 \\
&  68 & 83.1210 & -66.4484 & 0.64 & 0.40 & 291.0 &  10.8 &  107 \\
&  69 & 83.1274 & -66.4454 & 0.65 & 1.08 & 288.5 &   3.6 &  779 \\
&  70 & 83.1556 & -66.4563 & 0.65 & 0.56 & 286.7 &  23.9 &  210 \\
&  71 & 83.1486 & -66.4695 & 0.65 & 0.92 & 287.3 &  10.2 &  573 \\
&  72 & 83.0675 & -66.4148 & 0.65 & 0.45 & 288.3 &  53.6 &  139 \\
&  73 & 83.0422 & -66.4164 & 0.66 & 0.67 & 291.5 &  19.5 &  304 \\
&  74 & 83.1367 & -66.4617 & 0.66 & 0.39 & 290.2 &  48.3 &  102 \\
&  75 & 83.1172 & -66.4541 & 0.67 & 0.22 & 290.6 &   2.0 &   34 \\
&  76 & 83.1345 & -66.4592 & 0.67 & 0.56 & 289.0 &  46.2 &  219 \\
&  77 & 82.9924 & -66.4275 & 0.67 & 0.55 & 287.2 &  19.6 &  213 \\
&  78 & 83.0547 & -66.4403 & 0.70 & 1.05 & 286.5 &  22.4 &  801 \\
&  79 & 83.0932 & -66.4599 & 0.70 & 0.84 & 289.2 &  74.2 &  512 \\
&  80 & 83.0811 & -66.4449 & 0.71 & 0.36 & 286.9 &  23.0 &   94 \\
&  81 & 82.9530 & -66.4162 & 0.71 & 0.47 & 287.6 &  25.3 &  164 \\
&  82 & 82.9809 & -66.4057 & 0.72 & 0.21 & 287.0 &   9.3 &   34 \\
&  83 & 83.0644 & -66.4091 & 0.73 & 0.31 & 282.8 &   5.8 &   72 \\
&  84 & 83.0343 & -66.3870 & 0.73 & 0.66 & 286.3 &  17.1 &  334 \\
&  85 & 83.1387 & -66.4603 & 0.74 & 0.76 & 276.0 &  21.7 &  444 \\
& 86 & 82.9851 & -66.4055 & 0.79 & 0.35 & 289.3 &  31.4 &   98 \\
& 87 & 83.0436 & -66.4140 & 0.79 & 0.57 & 288.2 &  92.2 &  263 \\
& 88 & 83.1076 & -66.4858 & 0.79 & 1.02 & 289.5 &  59.1 &  864 \\
& 89 & 83.1251 & -66.4458 & 0.81 & 0.40 & 290.5 &   2.3 &  135 \\
& 90 & 82.9958 & -66.4127 & 0.82 & 0.43 & 286.7 &  25.1 &  155 \\
& 91 & 83.0812 & -66.4488 & 0.83 & 0.47 & 288.8 &  53.2 &  187 \\
& 92 & 83.0641 & -66.4442 & 0.84 & 0.41 & 287.1 &  42.2 &  148 \\
& 93 & 83.1064 & -66.4841 & 0.87 & 0.49 & 285.1 &  15.8 &  219 \\
& 94 & 83.0769 & -66.4368 & 0.87 & 0.48 & 290.1 &  32.6 &  211 \\
& 95 & 83.0286 & -66.4206 & 0.88 & 0.54 & 293.6 &  31.0 &  268 \\
& 96 & 83.0659 & -66.4510 & 0.88 & 0.59 & 289.6 &  12.9 &  324 \\
& 97 & 83.1716 & -66.4566 & 0.89 & 0.27 & 287.5 &   2.5 &   69 \\
& 98 & 83.0303 & -66.4136 & 0.89 & 0.54 & 285.6 &  17.8 &  272 \\
& 99 & 83.1193 & -66.4827 & 0.89 & 0.61 & 288.4 &  54.1 &  344 \\
& 100 & 82.9686 & -66.4181 & 0.90 & 0.46 & 284.8 &  45.7 &  196 \\
& 101 & 83.1452 & -66.4441 & 0.92 & 0.63 & 290.8 &  29.8 &  385 \\
& 102 & 83.1126 & -66.4494 & 0.92 & 0.44 & 290.0 &  17.7 &  182 \\
& 103 & 83.1090 & -66.4331 & 0.93 & 0.73 & 287.3 &  16.7 &  521 \\
& 104 & 83.0357 & -66.4396 & 0.94 & 0.72 & 288.0 &  10.1 &  504 \\
& 105 & 82.9541 & -66.4189 & 0.95 & 0.76 & 286.9 &  17.2 &  576 \\

\hline
\end{tabular}\\
\end{table*}

\addtocounter{table}{-1}

\begin{table*}
 \addtolength{\tabcolsep}{2pt}
%\addtocounter{table}
\centering
\caption{Continued}
\label{t_lines3}
\begin{tabular}{@{}ccccccccccccccccc}
\hline
%     & RA      &     Dec     &     Radius  &     Flux     &     Velocity dispersion & Luminosity& Virial mass \\
%              &             &             &              &          $\sigma$         & L$_{co}$     & M$_\odot$    \\
%\hline
& Number ID&  R.A.  & Decl. & Radius &$\sigma_v$ & Velocity  L$_{\rm co}$ & M$_{\rm VIR}$\\

&       &  deg   & dec& pc       & km\,s$^{-1}$          &   km\,s$^{-1}$&  K\,km\,s$^{-1}$\,pc$^2$      &  M$_{\odot}$ \\
\hline
& 106 & 83.0639 & -66.4578 & 0.95 & 0.69 & 288.3 &  56.6 &  473 \\
& 107 & 83.1418 & -66.4320 & 0.96 & 0.47 & 288.1 &  42.2 &  224 \\
& 108 & 82.9232 & -66.4267 & 0.96 & 0.60 & 283.7 &  34.9 &  362 \\
& 109 & 83.0884 & -66.4765 & 0.96 & 0.66 & 285.3 &   9.3 &  436 \\
& 110 & 82.9230 & -66.4328 & 0.98 & 0.45 & 283.7 &  14.2 &  202 \\
& 111 & 83.0581 & -66.4685 & 0.98 & 0.61 & 286.1 &  14.7 &  374 \\
& 112 & 83.1203 & -66.4719 & 1.03 & 0.38 & 290.6 &  64.3 &  151 \\
& 113 & 83.1401 & -66.4429 & 1.22 & 0.78 & 287.4 &  11.3 &  775 \\
& 114 & 83.0714 & -66.4043 & 1.24 & 0.69 & 283.2 & 123.6 &  608 \\
& 115 & 83.1155 & -66.4536 & 1.25 & 1.26 & 287.5 &  10.9 & 2068 \\
& 116 & 83.1556 & -66.4535 & 1.25 & 0.64 & 287.9 &   8.4 &  534 \\
& 117 & 83.1110 & -66.4432 & 1.33 & 1.12 & 288.6 &  10.2 & 1724 \\
& 118 & 83.1530 & -66.4663 & 1.55 & 0.78 & 290.3 &  10.7 &  977 \\
& 119 & 83.1245 & -66.4425 & 1.64 & 1.99 & 289.6 &  33.0 & 6775 \\
& 120 & 83.1702 & -66.4543 & 1.64 & 0.55 & 288.2 &  11.0 &  519 \\
& 121 & 83.1608 & -66.4537 & 1.99 & 0.68 & 290.4 &  16.6 &  952 \\

\hline
\end{tabular}\\
\end{table*}
%\section{Comparison with star formation}

%\begin{figure}
%\centering
%\includegraphics[scale=0.3]{PAH-12CO-yso-new.eps}
%\centering
%\caption{$^{12}$CO(1-0) emission in contour on IRAC %8.0${\,\rm \mu m}$ map for comparison. All identified %YSO's are shown in blue boxes and circles.}
%\label{yso}
%\end{figure}

%\section{Conclusions}

%keep your references in a bib file, such as reference.bib, which I made for you. It will automatically compile, if you use the natbib package and
%include the lines below. There is no need to include a file with \bibitems.

\section*{Acknowledgments}
This paper makes use of the following ALMA data:\\
\dataset[ADS/JAO.ALMA\#2013.1.00214.S]{https://almascience.nrao.edu/aq/project_code=2013.1.00214.S} and \dataset[ADS/JAO.ALMA\#2012.1.00335.S]{https://almascience.nrao.edu/aq/project_code=2012.1.00335.S}. ALMA is a partnership
of the ESO, NSF, NINS, NRC, NSC, and ASIAA. The Joint ALMA Observatory
is operated by the ESO, AUI/NRAO, and NAOJ. This work is financially
supported by NAOJ ALMA Scientific Research Grant Number 2016-03B.
F. Kemper acknowledges grant MOST104-2628-M-001-004-MY3 awarded by the
Ministry of Science and Technology in Taiwan. S. Hony acknowledges
financial support from DFG programme HO 5475/2-1. We thank Dario Colombo for providing the Python code for the bootstrap estimation of uncertainties.
%This research made use of \software{CASA, astrodendro (http://www.dendrograms.org/), Astropy (Astropy Collaboration, 2013)}.

\bibliographystyle{apj3}
{
\bibliography{N55_alma}
}

\end{document}